\documentclass[conference]{IEEEtran}
\IEEEoverridecommandlockouts
\usepackage{cite}
\usepackage{amsmath,amssymb,amsfonts}
\usepackage{algorithmic}
\usepackage{graphicx}
\usepackage{textcomp}
\usepackage{xcolor}

\usepackage{cite}

\usepackage{amsmath,amssymb,amsfonts}
\usepackage{url}
\usepackage[ruled,vlined,linesnumbered]{algorithm2e}

\usepackage{amsmath}
\usepackage{amsmath} % for \dfrac command
\usepackage{graphicx}
\usepackage{tikz}
\usepackage{adjustbox}
\usepackage{textcomp}
\usepackage{xcolor}
\usepackage{graphicx}
\usepackage{lettrine}
\usepackage{subfigure}
\usepackage{enumitem}
\usepackage{amsmath}
\usepackage{url} % This package is used to format URLs correctly
\usepackage{comment}
\definecolor{lightblue}{RGB}{173, 216, 230}
\usepackage{xcolor}
\usepackage{tabularx}
\usepackage{lipsum}
\usepackage{multirow}
\usepackage{multirow}
\usepackage{amsmath}
\usepackage{subcaption} % in preamble
\usepackage{subcaption}
\usepackage{graphicx}
\usepackage{soul}
\usepackage{xcolor}
\usepackage{caption}

\def\BibTeX{{\rm B\kern-.05em{\sc i\kern-.025em b}\kern-.08em
    T\kern-.1667em\lower.7ex\hbox{E}\kern-.125emX}}
\begin{document}

\title{A Multihop Rendezvous Protocol for Cognitive Radio-based Emergency Response Network\\}
\author{\IEEEauthorblockN{Zahid Ali$^{1}$, Saritha Unnikrishnan$^{2}$, Eoghan Furey$^{1}$, Ian McLoughlin$^{3}$ and Saim Ghafoor$^{1}$ }
	\IEEEauthorblockA{\textit{Department of Computing, Atlantic Technological University, (Donegal$^{1}$, Sligo$^{2}$ and Galway$^{3}$), Ireland} \\
		{Email: {\{L00177823, saritha.unnikrishnan, eoghan.furey, ian.mcloughlin, saim.ghafoor\}}@atu.ie}
	}
}

\title{A Multihop Rendezvous Protocol for Cognitive Radio-based Emergency Response Network\\
%{\footnotesize \textsuperscript{*}Note: Sub-titles are not captured for https://ieeexplore.ieee.org  andshould not be used}
\thanks{This Research has been supported by the Atlantic Technological University, Ireland under the Postgraduate Research Training Program in Modeling and Computation for Health and Society (MOCHAS PRTP).}
}

\maketitle

\begin{abstract}
This paper addresses the challenge of efficient rendezvous in multihop cognitive radio networks, where existing channel-hopping algorithms designed for single-hop scenarios incur increased delay and coordination inefficiencies in multi-node topologies. To overcome these limitations, we propose a Multihop Dual Modular Clock Algorithm (M-DMCA), which systematically extends modular clock-based rendezvous to multihop environments while preserving efficient channel coordination. The proposed scheme enables dual-channel selection per timeslot and incorporates a lightweight three-way handshake mechanism to improve coordination among intermediate nodes. Simulation results under worst-case conditions, including high primary user activity, asymmetric channel availability, and dense network settings, demonstrate that M-DMCA significantly reduces rendezvous time compared to existing approaches, achieving up to 24\% improvement. These results demonstrate the suitability of M-DMCA for timely node discovery in dynamic emergency response scenarios.
\end{abstract}

\section{Introduction}
\begin{figure}[ht]
	\centering
	\includegraphics[width=0.8\linewidth]{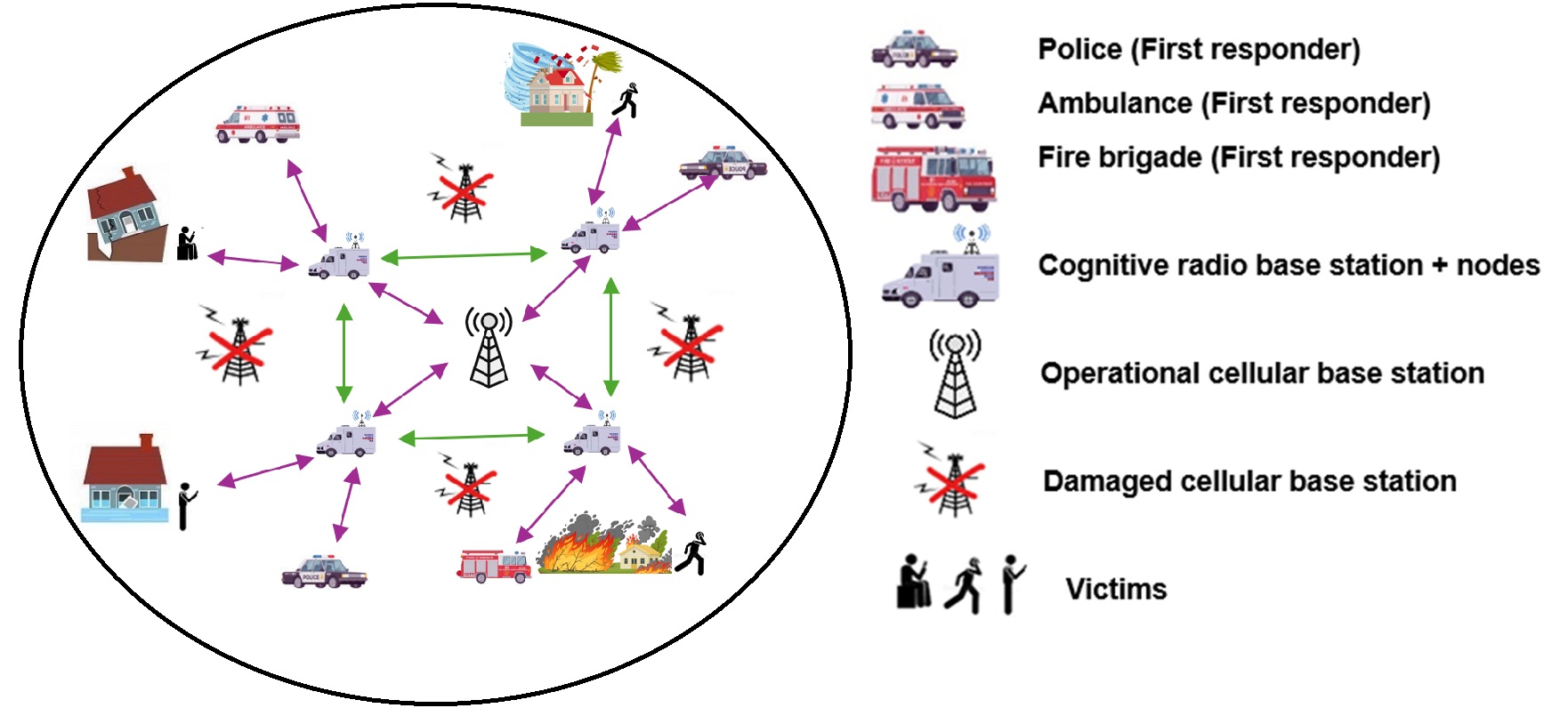} % Adjust the width as needed
	\caption{Example of CR based emergency response network}
	\label{fig:Example}
\end{figure}

\lettrine{I}{n} post-disaster scenarios, natural calamities such as earthquakes, floods, and hurricanes often destroy conventional communication infrastructure, critically hindering emergency response operations \cite{1,2}. Cognitive radio (CR) networks provide a promising solution by enabling rapidly deployable and autonomous communication systems that opportunistically access licensed spectrum \cite{4}. Such networks are particularly suitable for emergency environments where spectrum availability is dynamic and unpredictable. In CR-based emergency response networks, multihop rendezvous protocols play a crucial role in enabling distributed nodes to autonomously discover neighbouring nodes, establish communication links, and exchange information without centralised coordination. Efficient multihop rendezvous is essential for rapid topology formation, effective spectrum utilisation, and reliable connectivity, all of which are critical for timely disaster response and coordination among first responders, as illustrated in Fig.~\ref{fig:Example}. However, designing efficient rendezvous mechanisms in multihop environments is significantly more challenging than in single-hop scenarios. 
The presence of multiple intermediate nodes, asymmetric channel availability, and dynamic primary radio (PR) activity introduces coordination complexity and increases rendezvous delay. Existing rendezvous protocols \cite{5}, primarily designed for two-node scenarios, do not effectively address these challenges in multihop settings. In \cite{Ali2025} (our prior work), a dual modular clock-based rendezvous algorithm (DMCA) was proposed for single-hop networks. While effective in single-hop scenarios, DMCA does not address multihop coordination challenges, leaving a critical gap in distributed CR networks.

To address this gap, this paper proposes a Multihop Dual Modular Clock Algorithm (M-DMCA), which builds on modular clock-based rendezvous principles while specifically targeting multihop CR networks. The proposed approach enables dual-channel selection per timeslot, optimally distributed channel hopping, and incorporates a lightweight three-way handshake mechanism to improve coordination efficiency while accounting for PR activity.

The main contributions of this paper are as follows:
\begin{itemize}
\item	A multihop rendezvous mechanism based on dual modular clock principles that efficiently supports distributed node discovery in CR networks
\item	A dual channel selection strategy enabling two rendezvous attempts per timeslot to reduce discovery delay
\item	A lightweight three-way handshake mechanism that minimises coordination overhead among intermediate nodes
\item	Comprehensive performance evaluation demonstrating a significant reduction in rendezvous time under dynamic and worst-case network conditions
\end{itemize}
The rest of this paper is organised as follows. Section II presents the related work. Section III describes the system model and preliminaries. Section IV presents the proposed M-DMCA protocol. Section V provides performance evaluation results, and Section VI concludes the paper.

\section{Related Work}
Rendezvous in cognitive radio networks has been extensively studied, with early approaches primarily focusing on single-hop scenarios. Random Channel Selection (RCS) and Modular Clock Algorithm (MCA) \cite{5} are among the foundational techniques, where nodes independently select channels to achieve rendezvous. Modified versions of MCA {5} have also been proposed to improve performance under different channel conditions. However, these approaches are largely limited to two-node communication and do not scale efficiently to multihop environments.
To address multiuser and multihop scenarios, the Jump-Stay (JS) algorithm \cite{13} was introduced, providing guaranteed rendezvous through predefined channel-hopping sequences. While JS supports multiuser environments, its reliance on deterministic hopping patterns and lack of consideration for primary radio (PR) activity limit its effectiveness in dynamic and spectrum-constrained scenarios, such as disaster response networks.

In \cite{6}, the Extended Modular Clock Algorithm (EMCA) was proposed to support multihop rendezvous by combining clock-based channel selection with phased neighbour discovery. EMCA improves rendezvous performance compared to earlier approaches; however, its extended termination phase and reliance on a two-way handshake mechanism introduce additional communication overhead, leading to increased rendezvous delay in time-critical applications.

Other CR based multihop rendezvous protocols have been proposed with virtual channel coding \cite{asterjadhi2009}, predetermined sequence generation \cite{song2015, 15, 16} to improve scalability and coordination among multiple nodes. Nevertheless, common limitations of these approaches are the underutilization of available resources, as they typically allow only a single rendezvous attempt per timeslot. %Later on, multichannel opportunistic rendezvous (MOR) was proposed with two different types of packets such as unicast and broadcast\cite{feng2019}. However, the work was mainly focused to increased the throughput contain deterministic deployment of 5 nodes and 10 channels. This limitation significantly impacts performance in disaster where nodes deal with dense network, random topology and unknown PR activity.
This limitation significantly impacts performance in dense networks and under high PR activity.

Our prior work, the Dual Modular Clock Algorithm (DMCA) \cite{Ali2025}, which optimally distributed channels into prime and non-prime, enables dual rendezvous attempts per timeslot for single-hop networks, improving rendezvous efficiency. However, DMCA does not address the challenges of multihop coordination, motivating the proposed M-DMCA protocol.
Overall, existing approaches either lack efficient multihop support or incur significant coordination overhead, highlighting the need for a rendezvous mechanism that enables efficient multihop operation while maximising channel utilisation and minimising delay.
%}

\section{Preliminaries}
This section presents an overview of the network architecture, the PR activity model, the dual-channel selection mechanism, and the timeslot structure.
\subsection{Network Model}
The network is modeled as a connected graph of randomly deployed static CR nodes, where each node maintains at least one neighbor to ensure multihop connectivity for rendezvous and topology discovery. Each node is equipped with a single radio interface capable of spectrum sensing. However, due to varying locations, the available channels may differ across nodes. We consider a slotted system with synchronised timeslots, where the duration of one timeslot is assumed to be 1 second which is sufficient to exchange packets between a sender and receiver, denoted as T and T/2 \cite{rr, ss}. For spectrum sensing, we adopt an energy detection model, though cyclostationary feature detection may also be employed for primary user identification \cite{tt}. 

\subsection{Primary Radio Activity Model}
The PR activity model can be used to estimate and quantify spectrum occupancy in real environments \cite{uu}. For unknown PR activity, we adopt a memoryless continuous-time alternating ON/OFF Markov renewal model to simulate primary user traffic \cite{bk}. A key feature of this model is its ability to represent the duration for which a channel $i$ remains busy $T^i_\text{ON}$ or idle $T^i_\text{OFF}$ . The channel utilization $U^i$ denotes the proportion of time that channel $i$ is occupied by a PR, as defined in Eq. (\ref{eq:Ui}).
\begin{equation}\label{eq:Ui}
	U^i =  \frac {E[T^i_{\text{ON}}]}{ E[T^i_{\text{ON}}] + E[T^i_{\text{OFF}}]} = \frac{\lambda_Y}{\lambda_X+\lambda_Y}
\end{equation}
The probability that channel $i$ is in the ON or OFF state at time $t$ can be determined using the following formulas. These probabilities satisfy the condition $P_{ON}(t) + P_{OFF}(t) = 1$, indicating that the sum of Eq. (\ref*{eq:Pon}) and Eq. (\ref*{eq:Poff}) corresponds to the maximum channel utilization $U^i$.
\begin{equation}\label{eq:Pon}
	P_{ON}(t) =  \frac{\lambda_Y}{\lambda_X+\lambda_Y} - \frac{\lambda_Y}{\lambda_X+\lambda_Y}e^{-(\lambda_X+\lambda_Y)t}
\end{equation}
\begin{equation}\label{eq:Poff}
	P_{OFF}(t) =  \frac{\lambda_X}{\lambda_X+\lambda_Y} + \frac{\lambda_Y}{\lambda_X+\lambda_Y}e^{-(\lambda_X+\lambda_Y)t}
\end{equation}
Using the aforementioned formulas, channel rate parameters $\lambda_X$ and $\lambda_Y$ for zero, low, long, and high activity levels are generated \cite{Ali2025}. These parameters are combined to form a mixed PR activity model that captures the unpredictable primary user behavior typical of disaster scenarios, as summarised in Table~\ref{tabp}.
\vspace*{-2.0pt}
\begin{table*}[htbp]
	\caption{Channel Rate Parameters of Unknown Primary Radio Activity}
	\centering
	\resizebox{\textwidth}{!}{%
		%\Large
		\small
		\renewcommand{\arraystretch}{1.2} % Adjusts the row heigh
		\begin{tabular}{|c|c|c|c|c|c|c|c|c|c|c|c|c|c|c|c|c|c|c|c|c|c|}
			\hline
			\multicolumn{2}{|c|}{Channels}& CH-1 & CH-2 & CH-3 & CH-4 & CH-5 & CH-6 & CH-7 & CH-8 & CH-9 & CH-10 & CH-11 & CH-12 & CH-13 & CH-14 & CH-15 & CH-16 & CH-17 & CH-18 & CH-19 & CH-20 
			\\
			\hline
			\multicolumn{2}{|c|}{PR Activity}& Zero & Low & Long & High & Zero & Low & Long & High & Zero & Low & Long & High & Zero & Low & long & High & Zero & Low & Long & High 
			\\
			\hline
			\multirow{3}{*}{Mix } 
			& $\lambda_X$  
			& 1000 & 1.0 & 0.25 & 0.22 & 1000 & 1.36 & 0.21 & 0.22 & 1000 & 1.26 & 0.22 & 0.23 & 1000 & 1.26 & 0.21 & 0.21  &  1000 & 1.28 & 0.20 & 0.21 
			\\
			\cline{2-22}
			% Row 2
			& $\lambda_Y$  
			& 0 & 0.21 & 0.25 & 1.44 & 0 & 0.22 & 0.24 & 1.58 & 0 & 0.22 & 0.24 & 1.25 & 0 & 0.21 & 0.22 & 1.06 &  0 & 0.22 & 0.20 & 1.09
			\\
			\cline{2-22}
			% Row 3
			& $U^i$ 
			& 0 & 0.17 & 0.50 & 0.86 & 0 & 0.13 & 0.53 & 0.87 & 0 & 0.14 & 0.52 & 0.84 & 0 & 0.14 & 0.51 & 0.83 & 0 & 0.14 & 0.50 & 0.83 
			\\
			\hline					
		\end{tabular}%
	}
	\label{tabp}
\end{table*}

\subsection{Dual Channel Clock and Timeslot}
\begin{figure}[ht]
	\centering
	\includegraphics[width=0.6\linewidth]{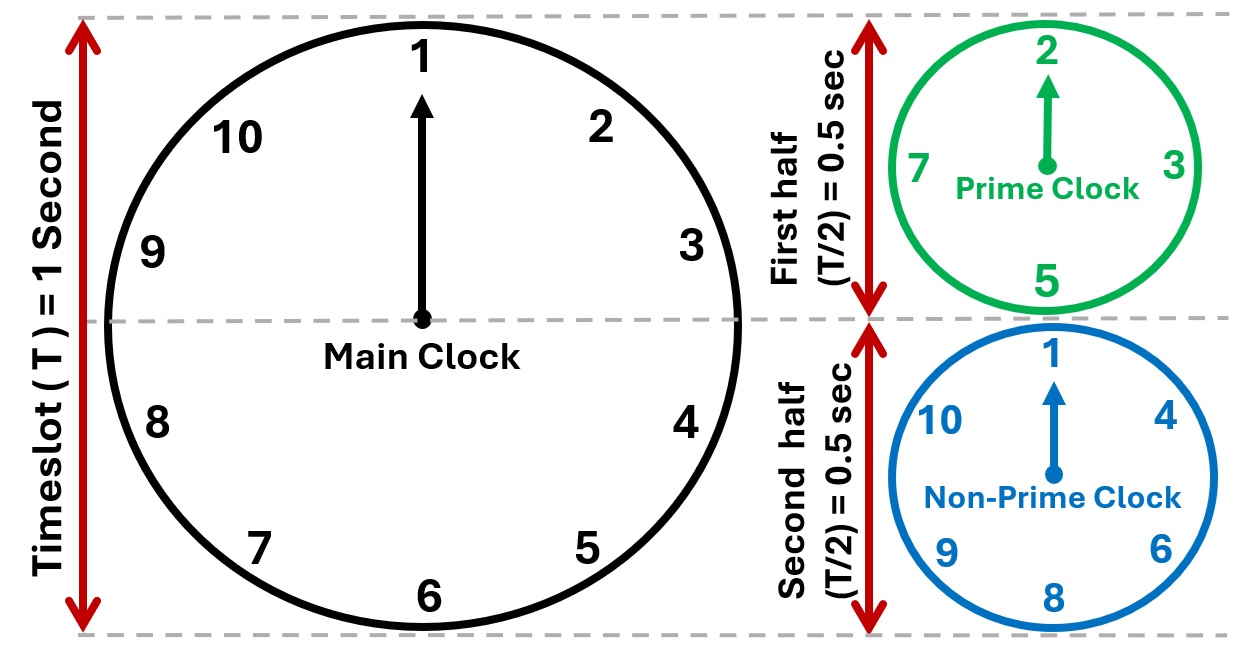} % Adjust the width as needed
	\caption{Distributed Clock Architecture and timeslot duration}
	\label{fig:DMCA_clock}
\end{figure}

To illustrate the dual-clock mechanism, we consider a channel set of 10 IDs for any node, denoted as $CU_{n}$= $\{1, 2, 3, 4, 5, 6, 7, 8, 9, 10\}$. Each node contains corresponding indices $\{0, 1, 2, 3, 4, 5, 6, 7, 8, 9\}$. Traditional random and clock-based algorithms, such as RCS, MCA, and EMCA, use these indices with a single main clock, as shown in Fig.~\ref{fig:DMCA_clock}. In contrast, the proposed M-DMCA first detects available channels and divides them into two sets, the prime channels $\{2, 3, 5, 7\}$ and non-prime channels $\{1, 4, 6, 8, 9, 10\}$, with indices hopping independently on each set. The timeslot duration remains 1 second for both single and dual channel selection. Since prime channels are fewer, the likelihood of achieving a rendezvous and completing a handshake is higher in the first half of the timeslot; if unsuccessful, the second half uses non-prime channels. In the unlikely event that either clock set is empty, the main clock is used.

\vspace*{-7.8pt}
\section{Multihop Dual Modular Clock Algorithm}
The multihop dual modular clock algorithm (M-DMCA) is an extention of DMCA \cite{Ali2025} which is based on number theory and modular clock algorithm \cite{5, 6}. The pseudocode of M-DMCA is presented in Algorithm 1, which considers the total number of nodes $N$ and channel information $m_i$. The multihop discovery process begins by identifying the available channel set $m_i$. This set is then divided into two subsets: prime channels $Mp$ and non-prime $Np$. Next, the algorithm determines the index lengths $i$ of $Mp$ and $Np$ to ensure successful discovery of all nodes using dual rendezvous and handshake. Each node must build a list of size $N-1$ that includes both its direct and indirect neighbours. For the first clock, node $i$ randomly selects an initial hoping index $j1^0$ and a hopping rate $R1$ for first half of timeslot. During this interval the node checks for presence of prime channel. If $Mp>0$, the node applies the modular function and directly hops within the prime channel set. Otherwise, it hops to the main channel set of size $m_i$. If no rendezvous occurs during the first half of the timeslot the node proceeds to the second clock. Node $i$ then randomly selects another initial hoping index $j2^0$ and a hopping rate $R2$ for second half of timeslot. In this phase, the node checks for the presence of non-prime channels. If $Np>0$, the node applies the modular function to hop directly within the non-prime channel set. Otherwise it hops to the main channel set of size $m_i$. This process continues until every node has discovered all other nodes $N-1$ through rendezvous, handshaking, and message transmission. However, before each transmission, every node must sense the status of PR activity and must not attempt rendezvous if the channel is occupied. The details of control messages that enable multihop rendezvous with exchange of information are provided below. 
\begin{enumerate}
	\item{Discovery Request (D-REQ)}: The process begins with the sender node broadcasting a neighbor discovery (D-REQ) message, which includes its Node ID along with a table of its directly and indirectly connected neighbors.
	\item {Discovery Response (D-RESP)}: When a node receives a D-REQ message, it updates its list of direct and indirect neighbors, and then sends back a D-RESP message containing its own Node ID along with the updated list.
	\item{Discovery Acknowledgment (D-ACK)}: Upon receiving a D-REQ in the case of two-way handshaking (2WH), or a D-RESP in the case of three-way handshaking (3WH), the node also confirms to the other node that its neighbour information has been updated according to the details in the packet.  
\end{enumerate}
\vspace*{-10.8pt}
\begin{algorithm}[ht]
	\smaller % Reduce font size for the algorithm content
	\caption{\mbox{Multihop Dual Modular Clock Algorithm}}
	\DontPrintSemicolon
	Input: $N$, total number of nodes \;
	Observed $m_i$, available channels \;
	Divide $m_i$ set into prime ($M_p$) and non-prime ($N_p$) sets \;
	Initialise index $j1_i^{0}$ [0,$m_i$)	randomly for first half \;
	Initialise index $j2_i^{0}$ [0,$m_i$)  randomly for second half \;
	%		Calculate the index length of $M_p$ and $N_p$ sets \;
	\While {node i not achieve N-1 with all nodes} {
		%		\If {\textbf{T} $>$ 0}{
			choose $R1$ from [0, $m_i$) randomly for first half \;
			choose $R2$ from [0, $m_i$) randomly for second half \;
			
			\For { {$t_i$= 0} to $m_i$ }{
				\textbf{{start rendezvous in first half of timeslot}\;}
				$j1_{\text{i}}^{t_i+1}$ = ($j1_{\text{i}}^{t_i}$ + $R1$) mod ($m_i$) \;
				
				\eIf {$M_p$ $>$ 0} { \textbf{c1} = $c_{\text{Mp, ($j1_{i}^{t_i+1}$) mod ($M_p$)}}$ // hop on prime }{c1 = $c_{\text{i, $j1_{i}^{t_i+1}$}}$ // hop on $m_i$ if $M_p=0$\;} 
				\enspace \textbf {sense} the PR Activity on $c1$ \;	    
				\enspace attempt a rendezvous only when $c1$ is not occupied\;
				% \enspace \eIf {channel c1 is occupied} {do not attempt rendezvous}
				% {attempt rendezvous on c1}
				\enspace \eIf {node i achieved N-1} {terminate rendezvous}
				{node i wait to complete first half of timeslot\;}
				%	    \textbf{end for}}

			\textbf{{start rendezvous in second half of timeslot}\;}

			%	\For { \textbf {$t$ = 0} to $m_i$ }{
				$j2_{\text{i}}^{t_i+1}$ = ($j2_{\text{i}}^{t_i}$ + $R2$) mod ($m_i$) \;
				
				\eIf {$N_p$ $>$ 0} {	
					\textbf{c2} = $c_{\text{Np, ($j2_{i}^{t_i+1}$) mod ($N_p$)}}$ // hop on non-prime
				}{ 
					c2 = $c_{\text{i, $j2_{i}^{t_i+1}$}}$ // hop on $m_i$ if $N_p=0$}
				\enspace \If {$(c2 = c1 )$ // only possible if $M_p$ or $N_p$ = 0\;} {    
					($j2_{\text{i}}^{t_i+1}$ + 1) mod ($m_i$) \;
					c2 = $c_{\text{i, $j2_{i}^{t_i+1}$}}$ }
				\enspace \textbf{sense} the PR Activity on $c2$ \; 
				\enspace attempt a rendezvous only when $c2$ is not occupied\;
				%\enspace \eIf{ channel c2 is occupied}{ do not attempt rendezvous}	
				%{attempt rendezvous on c2}
				\enspace \eIf {node i achieved N-1} {terminate rendezvous}
				{node i wait to complete second half of timeslot\;}
				\textbf {$t_i$ =  $t_i + 1$ \;}	
				\textbf{end for}}
			%	\textbf{(T+1)}} 
		\textbf{end while}}
\end{algorithm}
\vspace*{-11.8pt}
\subsection{Handshaking in Multihop Network}
In this work, we propose a three-way handshake (3WH) mechanism for multihop rendezvous, aimed at rapidly discovering the network topology with list of both direct neighbour list (DNL) and indirect neighbors list (INL) in minimal time.
\subsubsection{Two Way Handshake (2-WH)}
\begin{figure}[ht]
	\centering
	\includegraphics[width=0.86\linewidth]{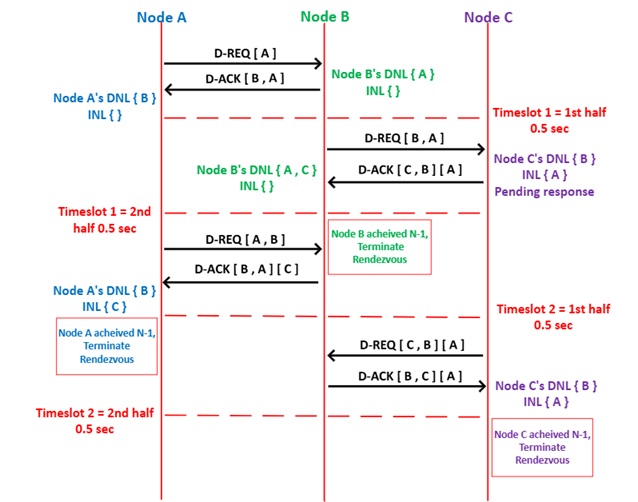} % Adjust the width as needed
	\caption{Two-way handshake (2-WH)}
	\label{fig:2WH}
\end{figure}

In the traditional 2WH approach \cite{6}, node A sends a D-REQ message with its current information. If received, node B responds with a D-ACK message, confirming the discovery and sharing its neighbor list. However, node B remains unaware about any updates or acknowledgment from node A until the next timeslot and rendezvous opportunity, which becomes inefficient in multihop scenarios as shown in  Fig.~\ref{fig:2WH}.
%\vspace*{-11.8pt}
\subsubsection{Three Way Handshake (3-WH)}
\begin{figure}[ht]
	\centering
	\includegraphics[width=0.86\linewidth]{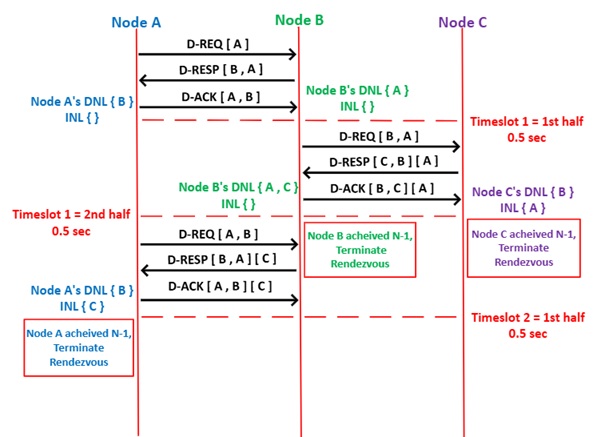} % Adjust the width as needed
	\caption{Three-way handshake (3-WH)}
	\label{fig:3WH}
\end{figure}

To address this limitation, 3WH adds an extra confirmation step. Node A sends a D-REQ message, and upon receiving it, node B replies with a D-RESP containing its neighbor information. Node A then transmits a final D-ACK message to confirm reception of the D-RESP and share its updated neighbor list. This ensures that both nodes are immediately synchronised with each other’s updated information, avoiding extra timeslots and rendezvous overhead in multihop settings as depicts in Fig.~\ref{fig:3WH}.

\section{Performance Evaluation}
We simulated the proposed M-DMCA multihop rendezvous protocol using two half-slots within a single time slot and implemented both 2WH and 3WH mechanisms in the NS-3.35 simulator. The implementation models the behaviour of a CR network, including MAC, dual-channel operation, handshake mechanisms, and PR activity \cite{chigan,6}. 
For comparison, we also implemented clock-based traditional RCS, MCA \cite{5}, and EMCA \cite{6}. Since these protocols were originally designed for a single rendezvous attempt per timeslot, we adapted them to support multihop and two rendezvous attempts per timeslot (2RATS) for fairness, aligning with the dual-rendezvous nature of M-DMCA.The results are shown in logarithmic (log) scale. 

 %The results are shown with 95\% confidence interval.
%\vspace*{-12.8pt}
\subsection{Simulation Setup}
For comparative analysis, we have considered static wireless topology where 3, 10 and 20 nodes are randomly distributed within area of 1000 $\times$ 1000$~\text{m}^2$. Each node has single radio interface with uniform transmission range of 100 meters. Random topologies are generated via the BonnMotion tool\cite{bonnmotion_web}. We consider both symmetric and asymmetric channel sets, where all nodes first sense PR activity before each rendezvous attempt. We compared the existing 2-WH and the proposed 3-WH multihop rendezvous protocols for discovering all connected neighbours. To asses the outcomes, we measured the Time to Rendezvous (TTR), defined as the duration from the start of the rendezvous process until all nodes are discovered within any half of a timeslot. The Average TTR (ATTR) is defined as the mean TTR value, first averaged across all nodes $(N)$ in a single simulation run, and then averaged across all simulation runs $(R)$, as expressed by~\eqref{eq:ADT}.
\begin{equation}
	\ ATTR =  \dfrac {1}{R} \sum_{j=1}^{R} \left(\dfrac {1}{N} \sum_{i=1}^{N} {TTR_i }\right)
	\label{eq:ADT}
\end{equation}

In addition, we define a metric termed Packets per Successful Rendezvous (PPR) to quantify the communication overhead associated with neighbour discovery. This metric represents the total number of packet transmissions required to achieve one successful rendezvous.
%For a single simulation run $j$, PPR is defined as:
Similar to ATTR, the average PPR over R simulation runs is computed as:

\begin{equation}
	\ PPR_{Avg} =  \dfrac {1}{R} \sum_{j=1}^{R} \left(
	\dfrac {PT_j }{SR_j}\right)
	\label{eq:PPRavg}
\end{equation}
%\begin{equation}
%	PPR_{j} =  \dfrac {PT_j }{SR_j}
%	\label{eq:PPR}
%\end{equation}
where $PT_j$ denotes the total number of packet transmissions in the jth simulation run, including control packets and retransmissions, and $SR_j$ represents the total number of successful rendezvous events. 
Lower values of PPR indicate higher efficiency, as fewer transmissions are required per successful rendezvous.

%\vspace*{-1.8pt}
\subsection{Impact of handshaking mechanisms (2WH vs 3WH)}
\begin{figure}[htbp]
	\centering
	% --- Left image (a) ---
	\begin{minipage}[b]{0.46\columnwidth}
		\centering
		\begin{tikzpicture}
			\node[draw, line width=0.4pt, inner sep=2pt] {%
				\includegraphics[width=1.36in]{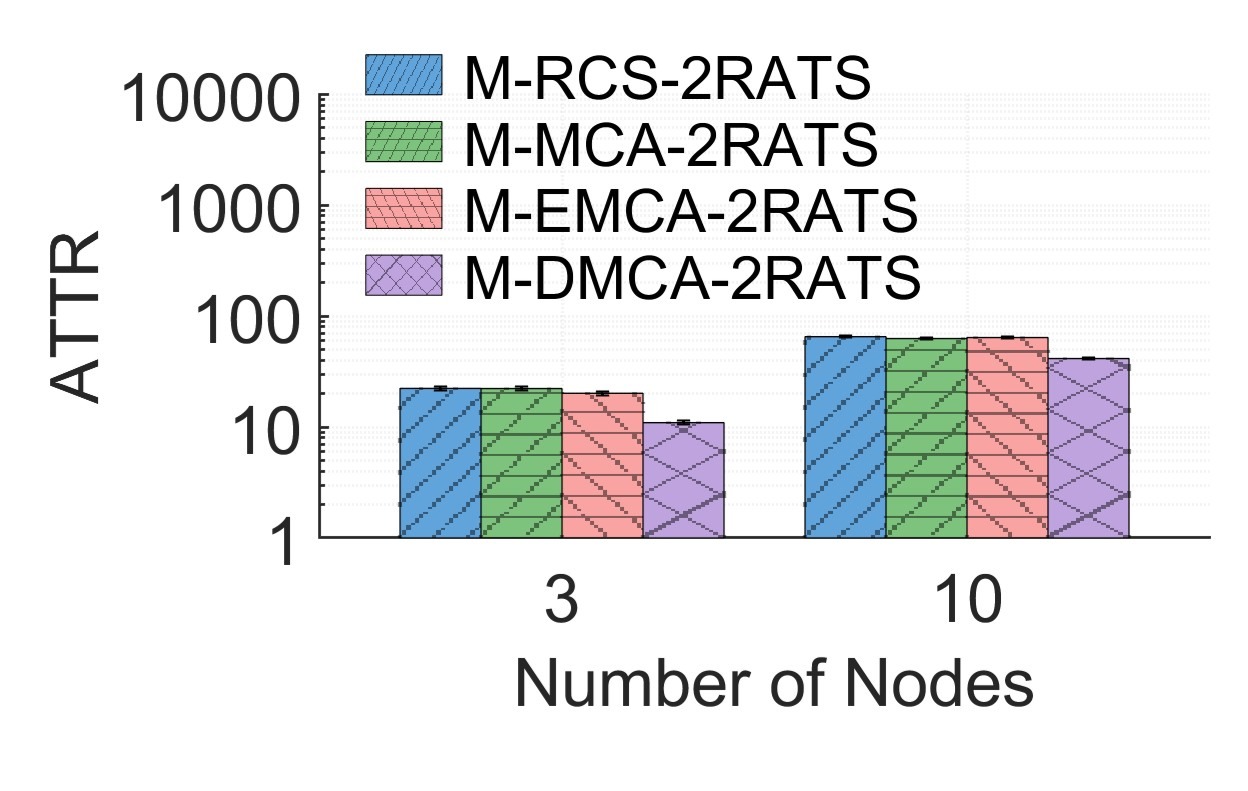} %	[width=2.5in]
			};
		\end{tikzpicture}\\[-0.25em]
		{\footnotesize (a)}
	\end{minipage}%
	% --- Right image (b) ---
	\begin{minipage}[b]{0.46\columnwidth}
		\centering
		\begin{tikzpicture}
			\node[draw, line width=0.4pt, inner sep=2pt] {%
				\includegraphics [width=1.36in]{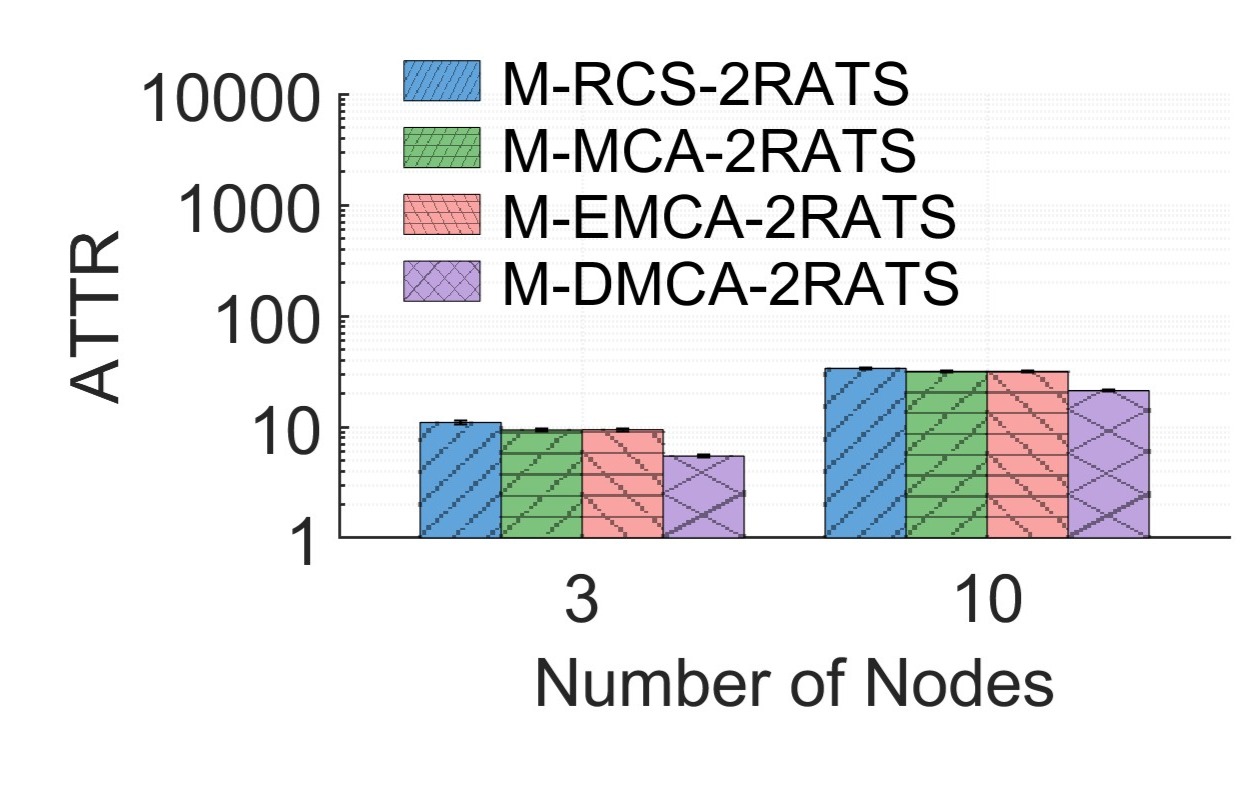}%[width=\linewidth] [width=2.5in]
			};
		\end{tikzpicture}\\[-0.25em]
		{\footnotesize (b)}
	\end{minipage}	
	\caption{Symmetric 10-CH with 0\%PR (a) 2-WH vs (b) 3-WH}\hfill
	\label{fig:R1}
\end{figure}

\begin{figure}[htbp]
	\centering	
	% --- Left image (a) ---
	\begin{minipage}[b]{0.46\columnwidth}
		\centering
		\begin{tikzpicture}
			\node[draw, line width=0.5pt, inner sep=2pt] {%
				\includegraphics[width=1.36in]{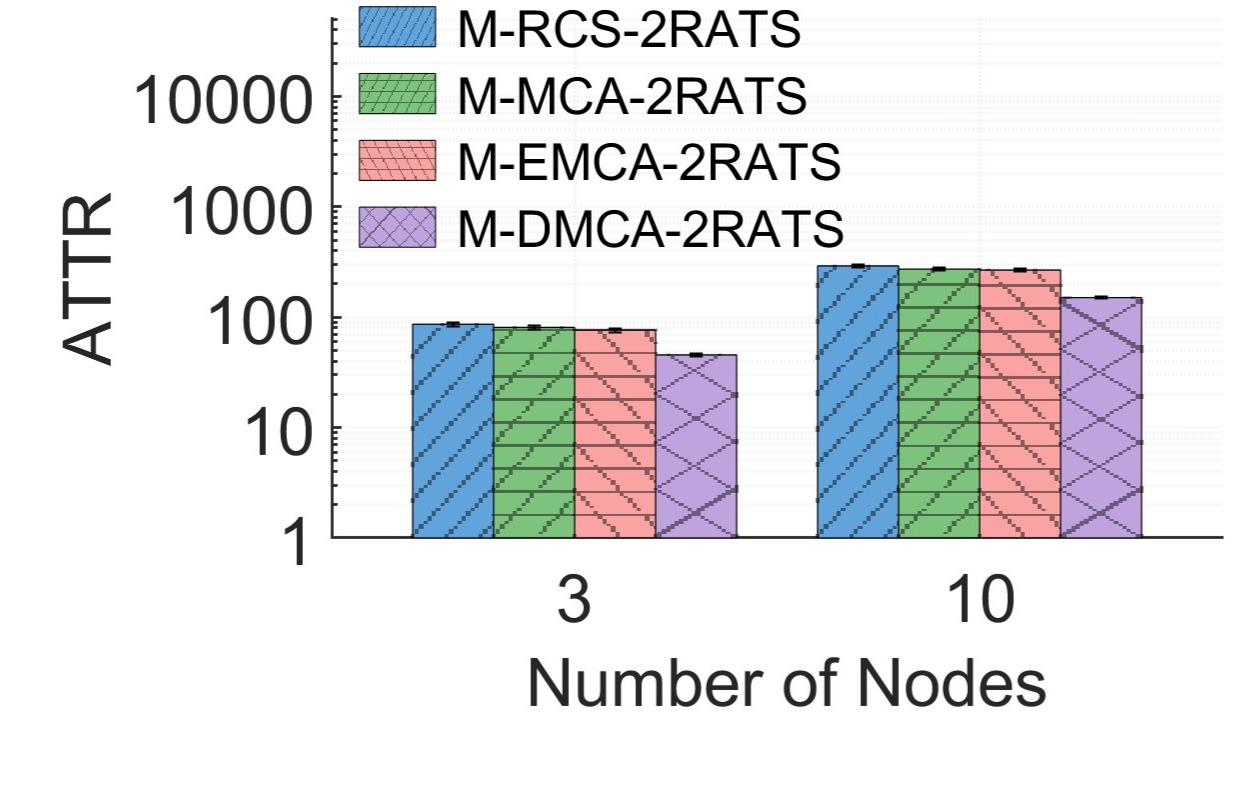}
			};
		\end{tikzpicture}\\[-0.25em]
		{\footnotesize (a)}
	\end{minipage}%
	% --- Right image (b) ---
	\begin{minipage}[b]{0.46\columnwidth}
		\centering
		\begin{tikzpicture}
			\node[draw, line width=0.5pt, inner sep=2pt] {%
				\includegraphics[width=1.36in]{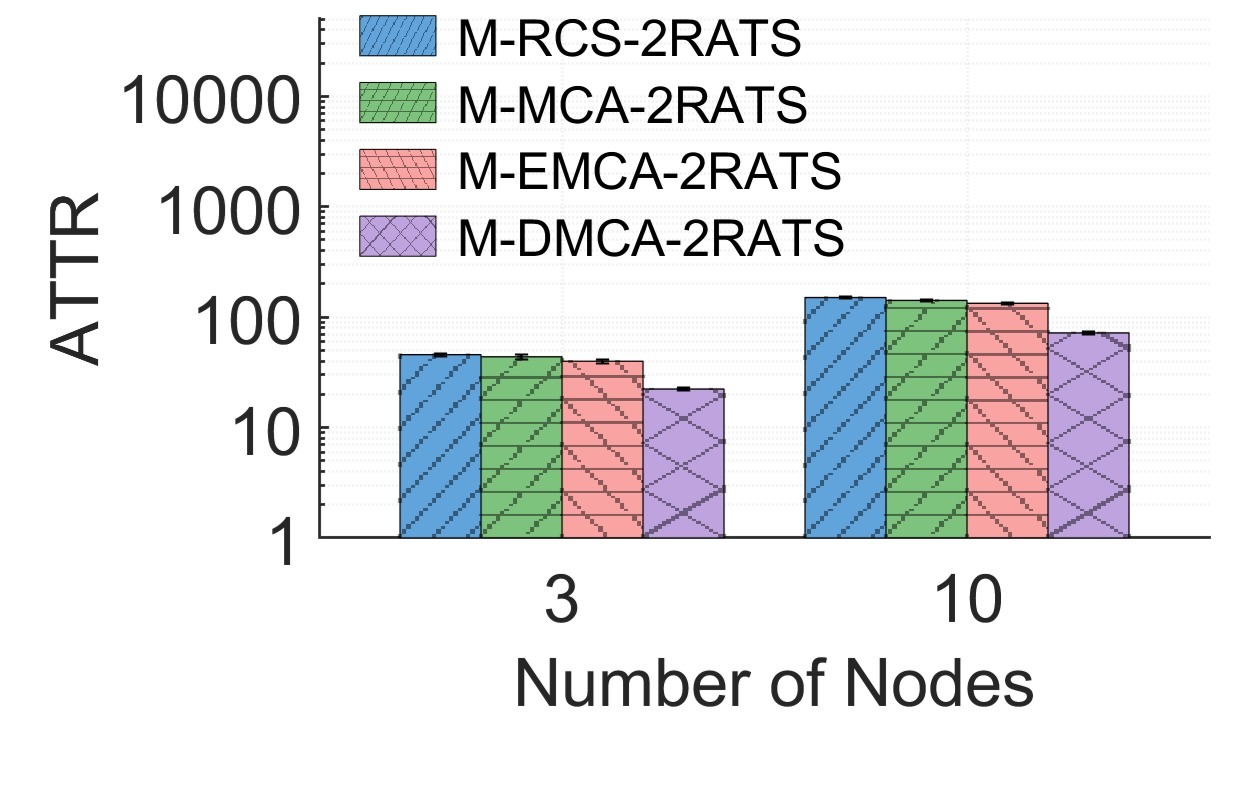}%
			};
		\end{tikzpicture}\\[-0.25em]
		{\footnotesize (b)}
	\end{minipage}
	\caption{Symmetric 10-CH with 85\%PR (a) 2-WH vs (b) 3-WH}\hfill
	\label{fig:R2}
\end{figure}

Fig.~\ref{fig:R1}, compares the ATTR for 2WH and 3WH under symmetric channel settings (10 channels) for 3 and 10 nodes under 0\% PR activity. The 3 WH improves the ATTR by almost 50\% compared to 2WH (Fig.~\ref{fig:R1}a and Fig.~\ref{fig:R1}b). The 3WH consistently achieves lower ATTR, particularly under high PR activity (85\%), shown in Fig.~\ref{fig:R2}a and Fig.~\ref{fig:R2}b. The proposed protocol M-DMCA also achieved improved ATTR of almost 33\% (0\% PR activity) and 49\% (85\% PR activity) compared to other multihop protocols. This improvement is attributed to enhanced confirmation reliability and an optimized prime channel list, which accelerates the rendezvous process. This becomes especially critical in multihop environments where channel availability is dynamic.  
\vspace*{-20.8pt}

\begin{figure}[ht]
	\centering
	% --- Left image (a) ---
	\begin{minipage}[b]{0.46\columnwidth}
		\centering
		\begin{tikzpicture}
			\node[draw, line width=0.4pt, inner sep=2pt] {%
				\includegraphics[width=1.36in]{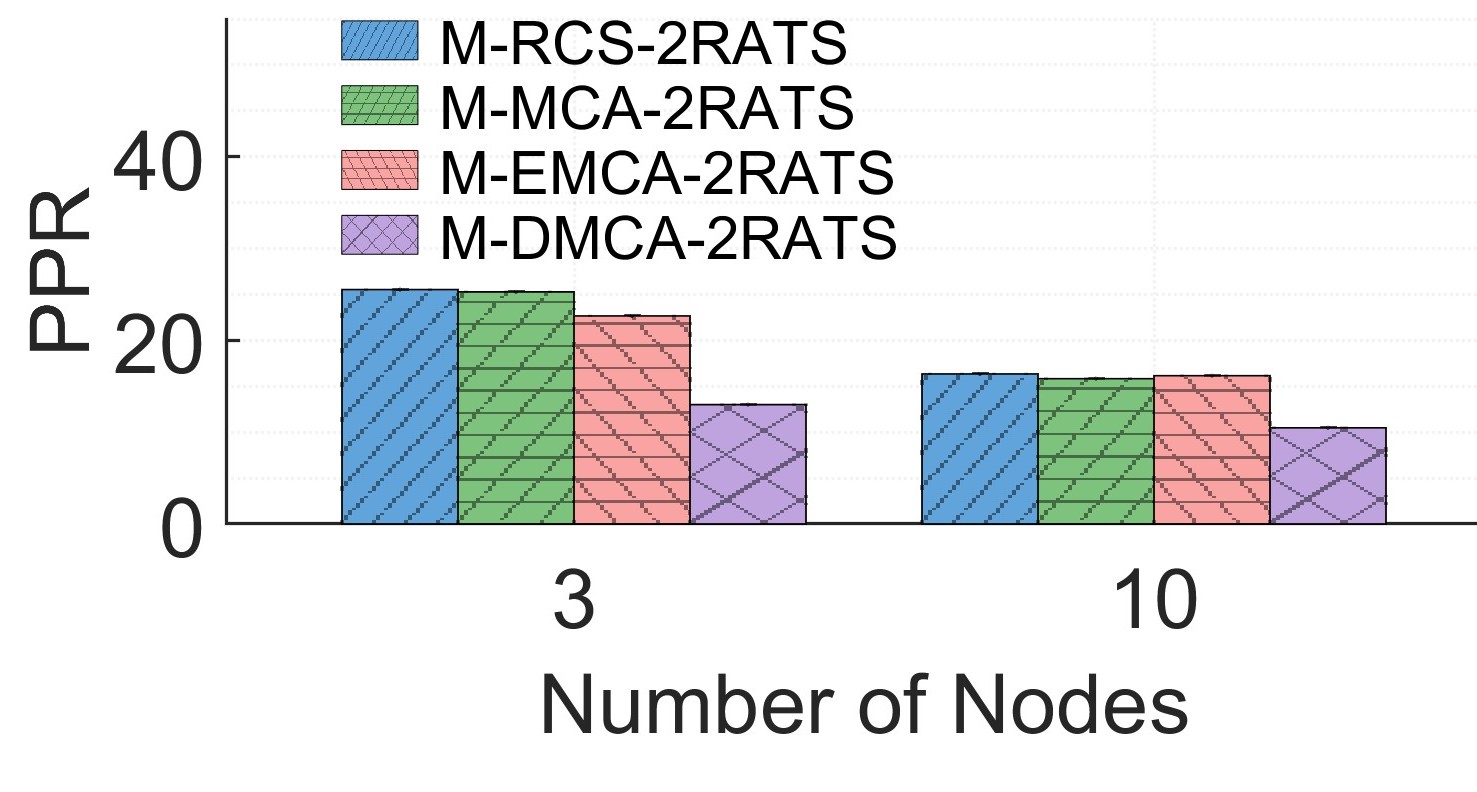} %	[width=2.5in]
			};
		\end{tikzpicture}\\[-0.25em]
		{\footnotesize (a)}
	\end{minipage}%
	% --- Right image (b) ---
	\begin{minipage}[b]{0.46\columnwidth}
		\centering
		\begin{tikzpicture}
			\node[draw, line width=0.4pt, inner sep=2pt] {%
				\includegraphics [width=1.36in]{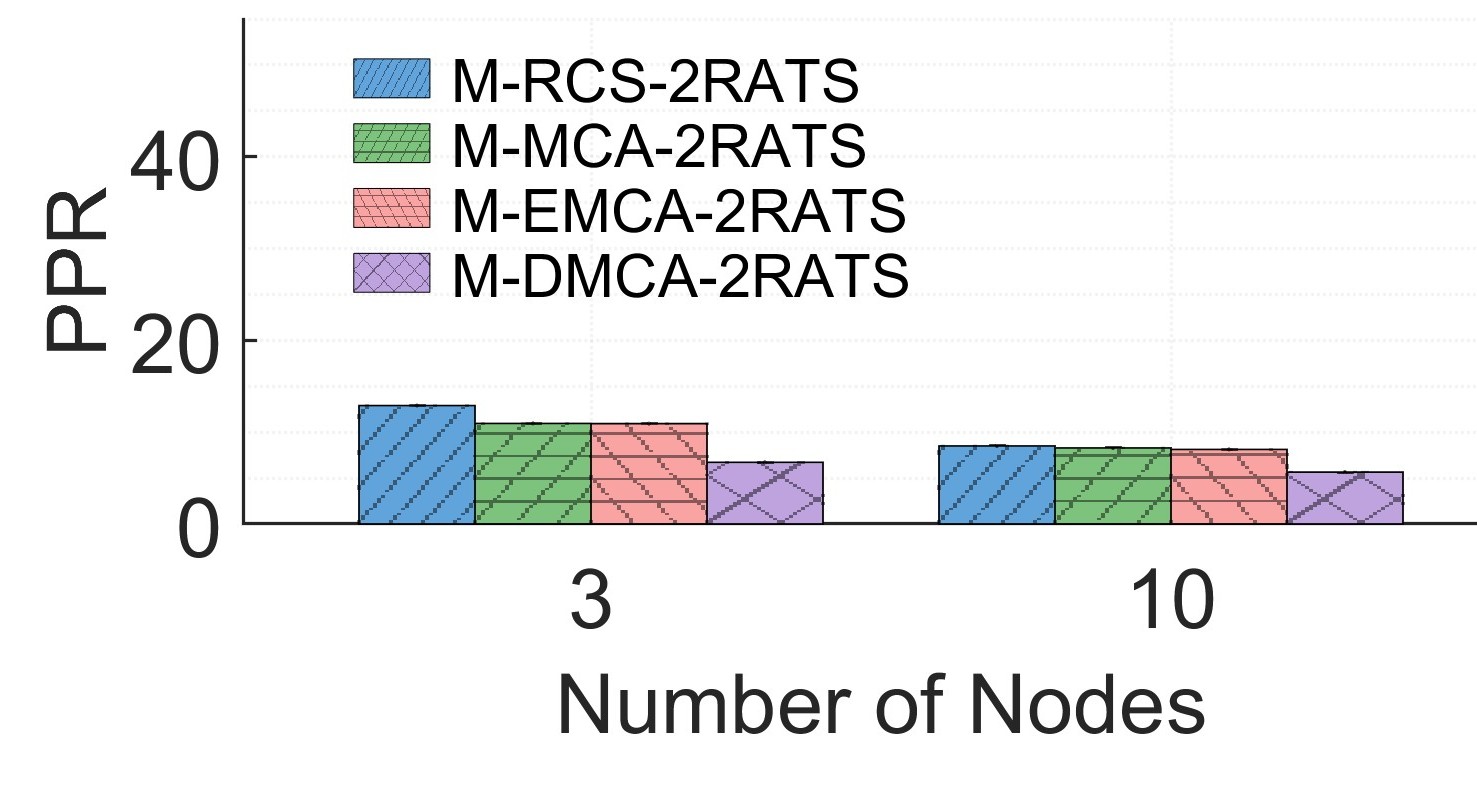}%[width=\linewidth] [width=2.5in]
			};
		\end{tikzpicture}\\[-0.25em]
		{\footnotesize (b)}
	\end{minipage}	
	\caption{Symmetric 10-CH with 0\%PR (a) 2-WH vs (b) 3-WH}\hfill
	\label{fig:hR1}
\end{figure}

\begin{figure}[ht]
	\centering	
	% --- Left image (a) ---
	\begin{minipage}[b]{0.46\columnwidth}
		\centering
		\begin{tikzpicture}
			\node[draw, line width=0.5pt, inner sep=2pt] {%
				\includegraphics[width=1.36in]{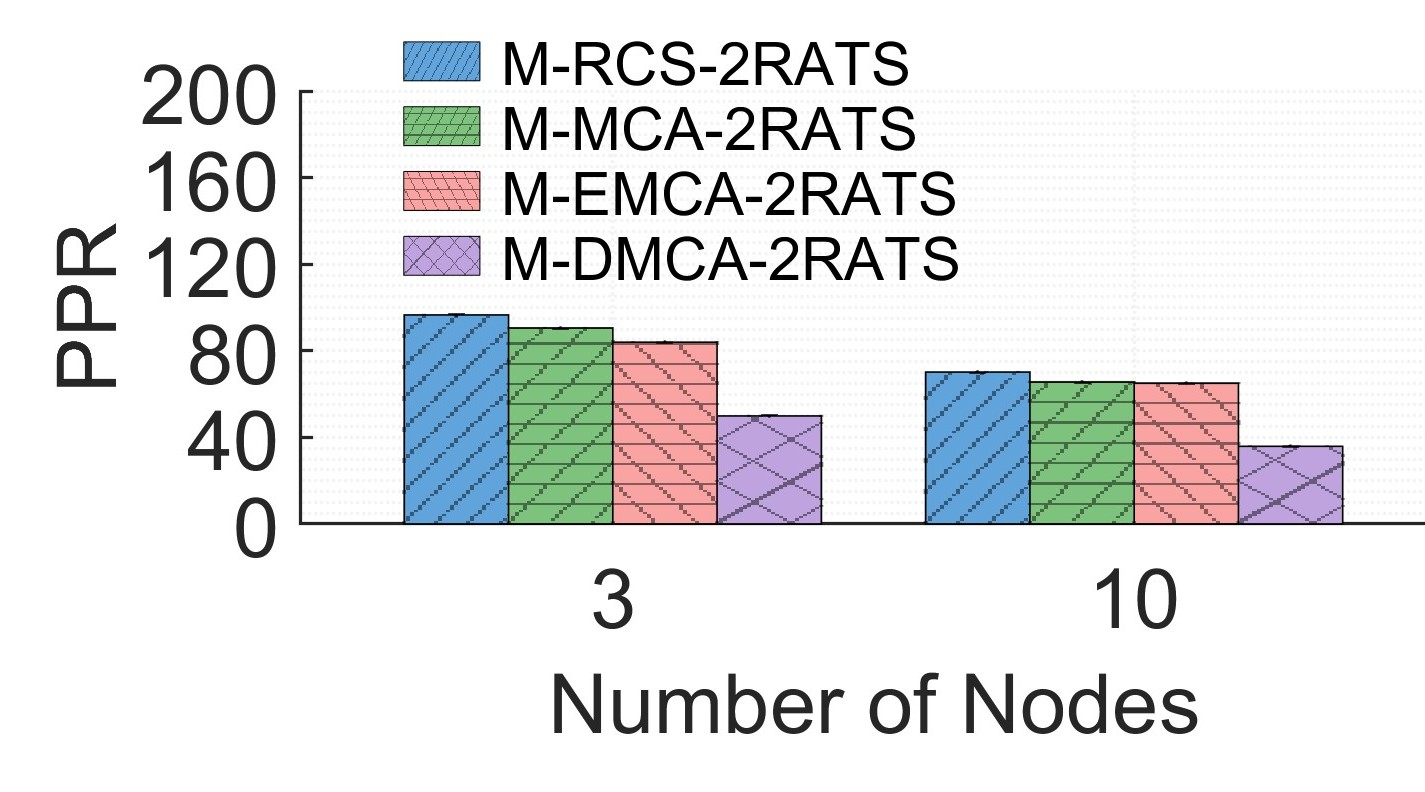}
			};
		\end{tikzpicture}\\[-0.25em]
		{\footnotesize (a)}
	\end{minipage}%
	% --- Right image (b) ---
	\begin{minipage}[b]{0.46\columnwidth}
		\centering
		\begin{tikzpicture}
			\node[draw, line width=0.5pt, inner sep=2pt] {%
				\includegraphics[width=1.36in]{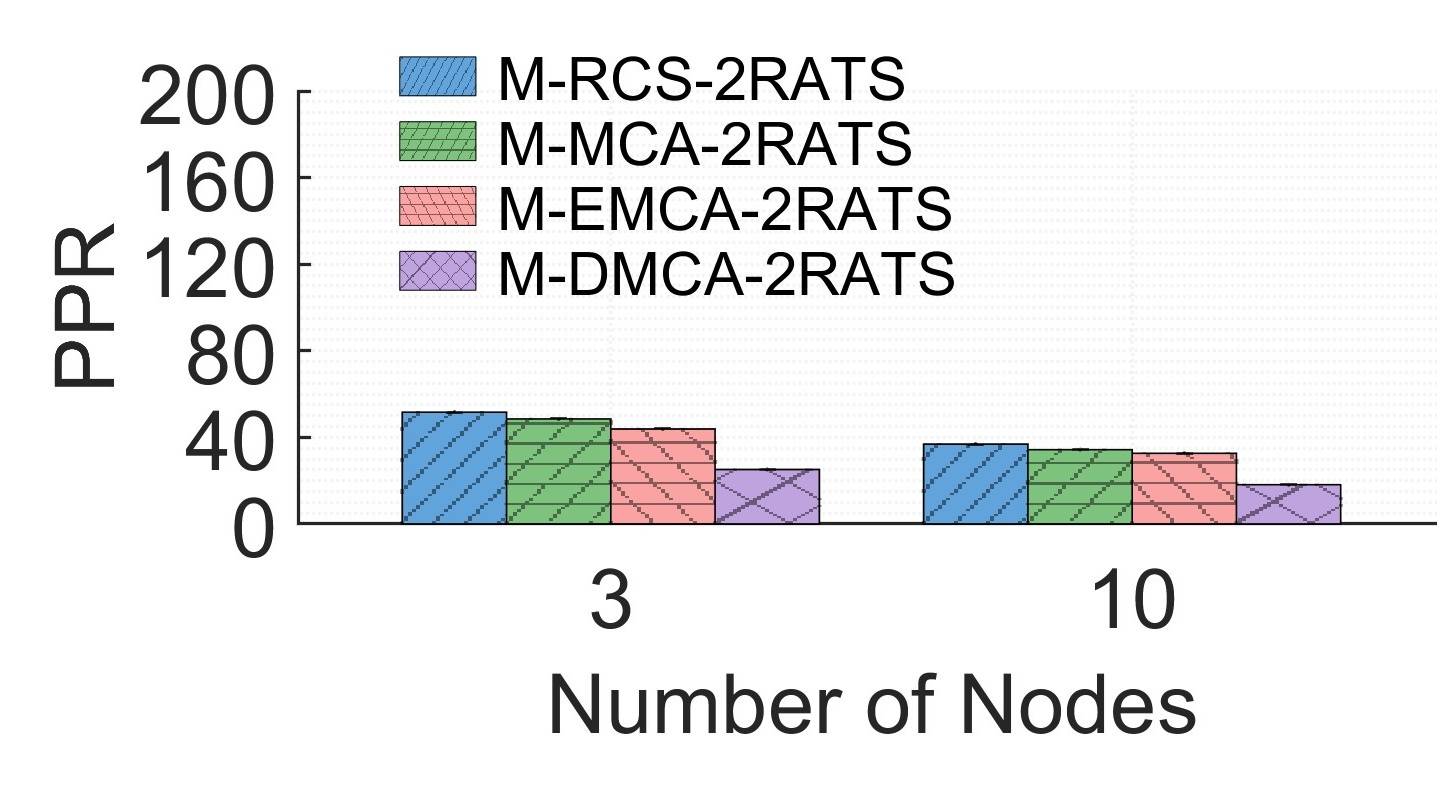}%
			};
		\end{tikzpicture}\\[-0.25em]
		{\footnotesize (b)}
	\end{minipage}
	\caption{Symmetric 10-CH with 85\%PR (a) 2-WH vs (b) 3-WH}\hfill
	\label{fig:hR2}
\end{figure}

While ATTR evaluates the time efficiency of rendezvous, PPR measures the communication overhead, quantifying the number of packet transmissions required per successful rendezvous and reflecting protocol efficiency in practical deployment. Fig.~\ref{fig:hR1} and Fig.~\ref{fig:hR2} illustrate the PPR for 2WH and 3WH under symmetric channel settings (10 channels) for 3 and 10 nodes, considering both 0\% and 85\% PR activity. The 3WH consistently achieves lower PPR than 2WH, indicating reduced communication overhead per successful rendezvous. Although a single 3WH handshake involves one extra packet compared to 2WH, it completes bidirectional confirmation in a single cycle, whereas 2WH often requires multiple cycles to achieve full confirmation. This reduces the number of repeated handshake attempts and retransmissions per successful rendezvous. Additionally, the PPR decreases as the number of nodes increases from 3 to 10, due to higher local connectivity and more rendezvous opportunities in denser topologies. The proposed M-DMCA protocol also demonstrates lower PPR compared to other multihop protocols, highlighting its ability to balance reliability and communication efficiency effectively, even in the presence of PR activity.

\vspace*{-14.8pt}
\vspace*{-14.8pt}
\subsection{Impact of channel similarity ratio (m)}
\begin{figure}[ht]
	\centering	
	% --- Left image (a) ---
	\begin{minipage}[b]{0.46\columnwidth}
		\centering
		\begin{tikzpicture}
			\node[draw, line width=0.5pt, inner sep=2pt] {%
				\includegraphics[width=1.36in]{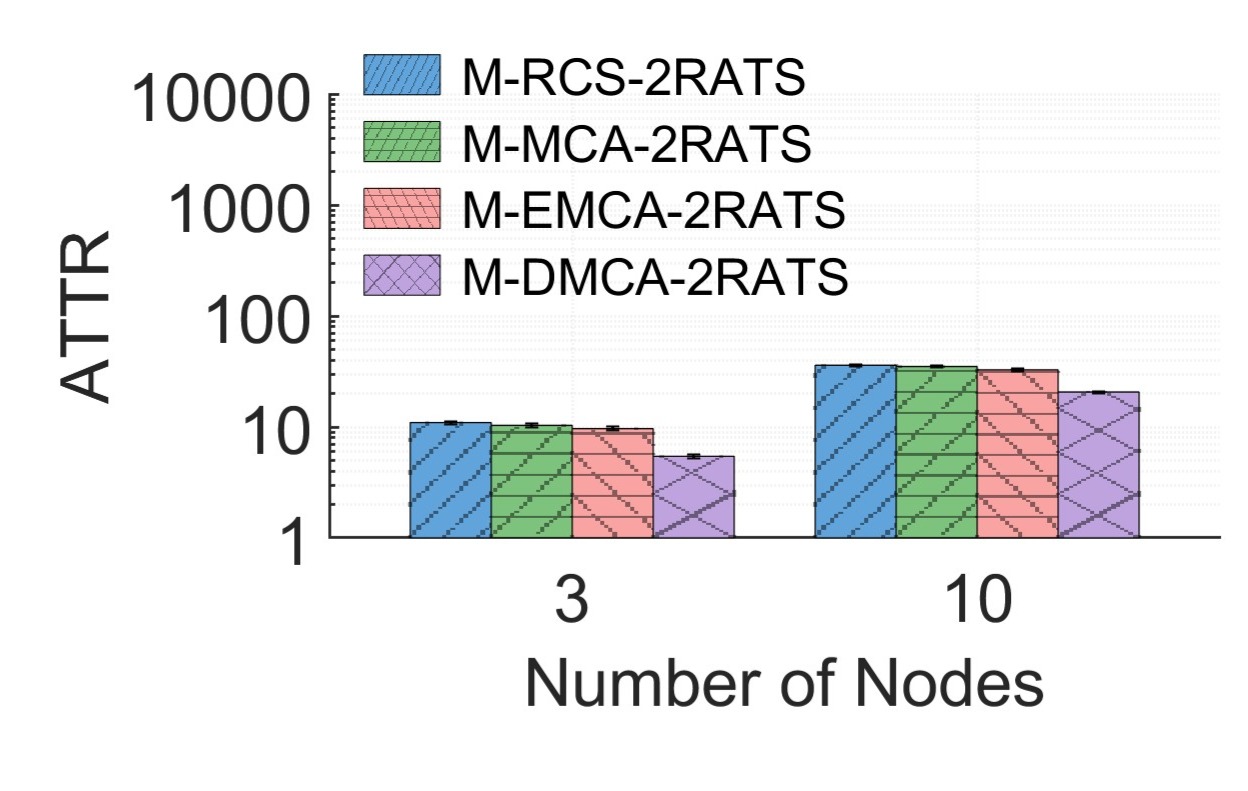}
			};
		\end{tikzpicture}\\[-0.25em]
		{\footnotesize (a)}
	\end{minipage}%
	% --- Right image (b) ---
	\begin{minipage}[b]{0.46\columnwidth}
		\centering
		\begin{tikzpicture}
			\node[draw, line width=0.5pt, inner sep=2pt] {%
				\includegraphics[width=1.36in]{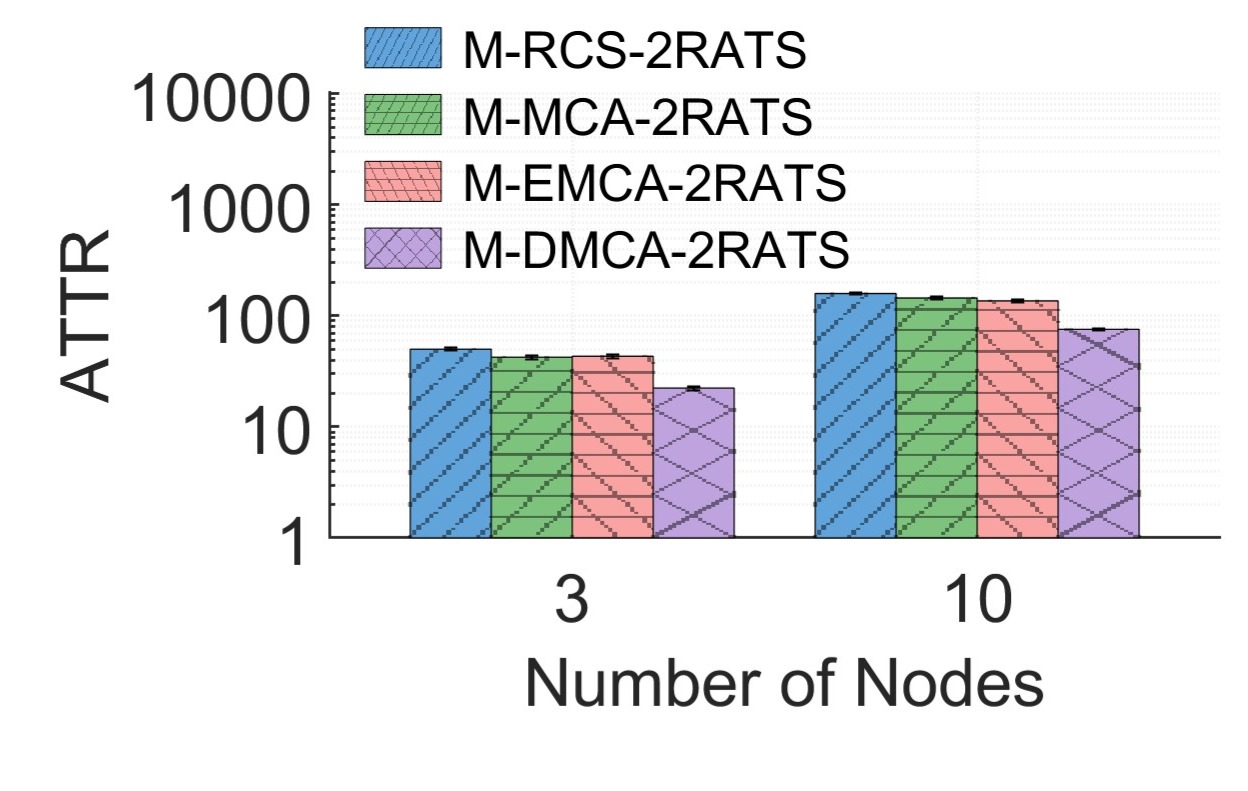}%
			};
		\end{tikzpicture}\\[-0.25em]
		{\footnotesize (b)}
	\end{minipage}
	
	\caption{Asym 10-CH with   m=9 (a) 0\%PR  and (b) 85\%PR}\hfill
	\label{fig:R6}
\end{figure}

\begin{figure}[ht]
	\centering	
	% --- Left image (a) ---
	\begin{minipage}[b]{0.46\columnwidth}
		\centering
		\begin{tikzpicture}
			\node[draw, line width=0.5pt, inner sep=2pt] {%
				\includegraphics[width=1.36in]{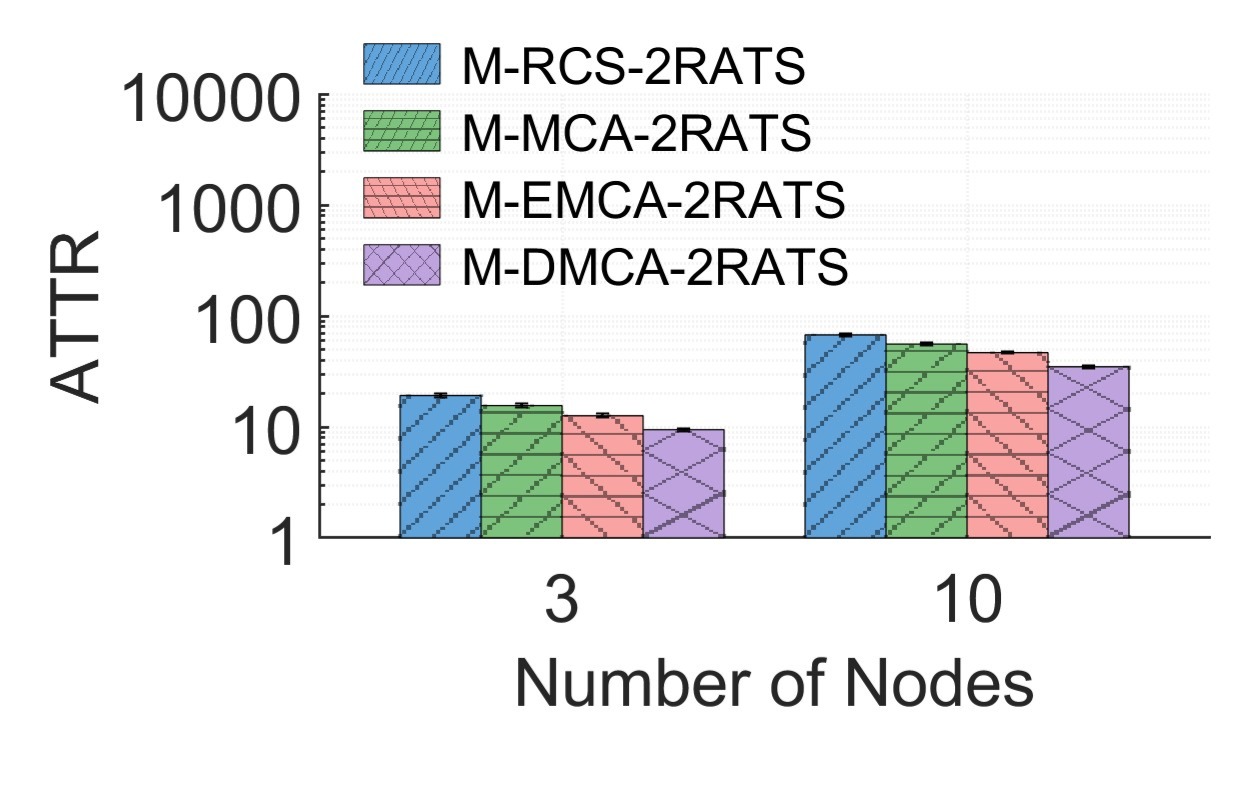}
			};
		\end{tikzpicture}\\[-0.25em]
		{\footnotesize (a)}
	\end{minipage}%
	% --- Right image (b) ---
	\begin{minipage}[b]{0.46\columnwidth}
		\centering
		\begin{tikzpicture}
			\node[draw, line width=0.5pt, inner sep=2pt] {%
				\includegraphics[width=1.36in]{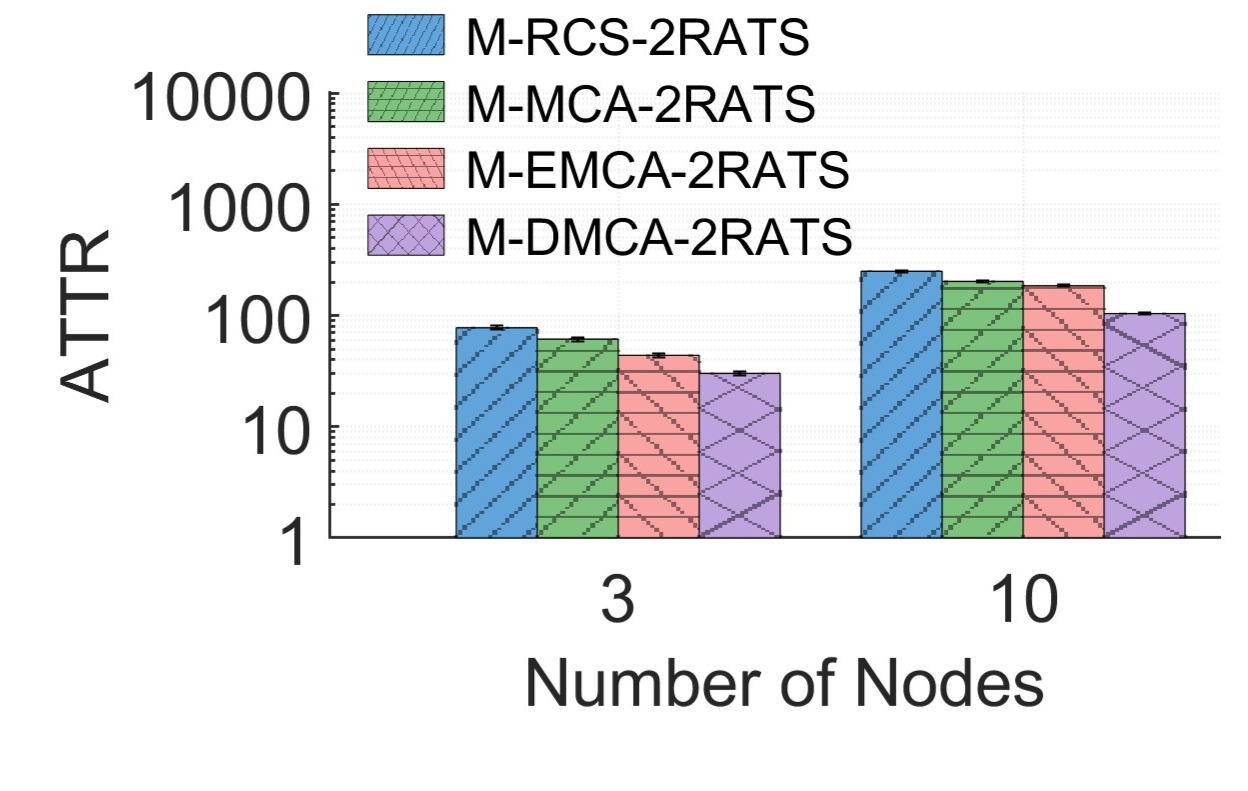}%
			};
		\end{tikzpicture}\\[-0.25em]
		{\footnotesize (b)}
	\end{minipage}
	\caption{Asym 10-CH with m=5  (a) 0\%PR  and (b) 85\%PR}\hfill
	\label{fig:R5}
\end{figure}

\begin{figure}[ht]
	\centering	
	% --- Left image (a) ---
	\begin{minipage}[b]{0.46\columnwidth}
		\centering
		\begin{tikzpicture}
			\node[draw, line width=0.5pt, inner sep=2pt] {%
				\includegraphics[width=1.36in]{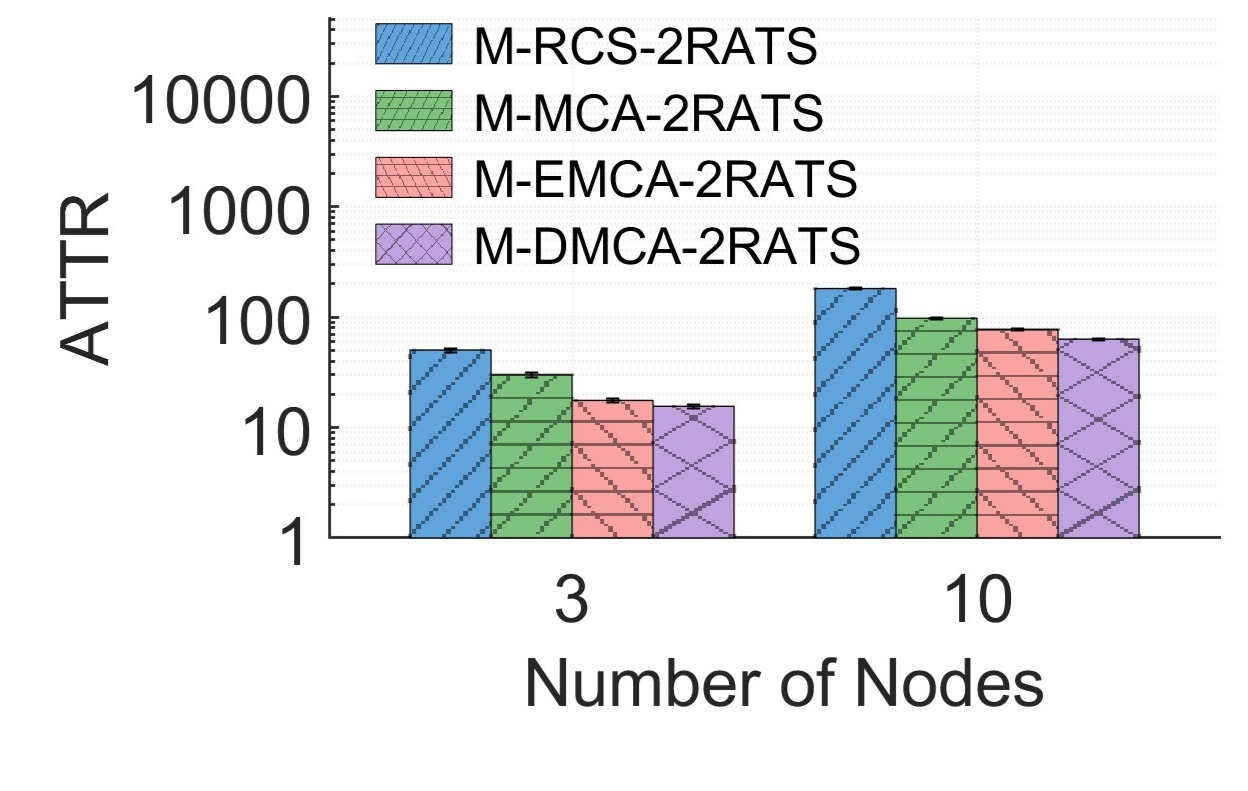}
			};
		\end{tikzpicture}\\[-0.25em]
		{\footnotesize (a)}
	\end{minipage}%
	%\hfill
	% --- Right image (b) ---
	\begin{minipage}[b]{0.46\columnwidth}
		\centering
		\begin{tikzpicture}
			\node[draw, line width=0.5pt, inner sep=2pt] {%
				\includegraphics[width=1.36in]{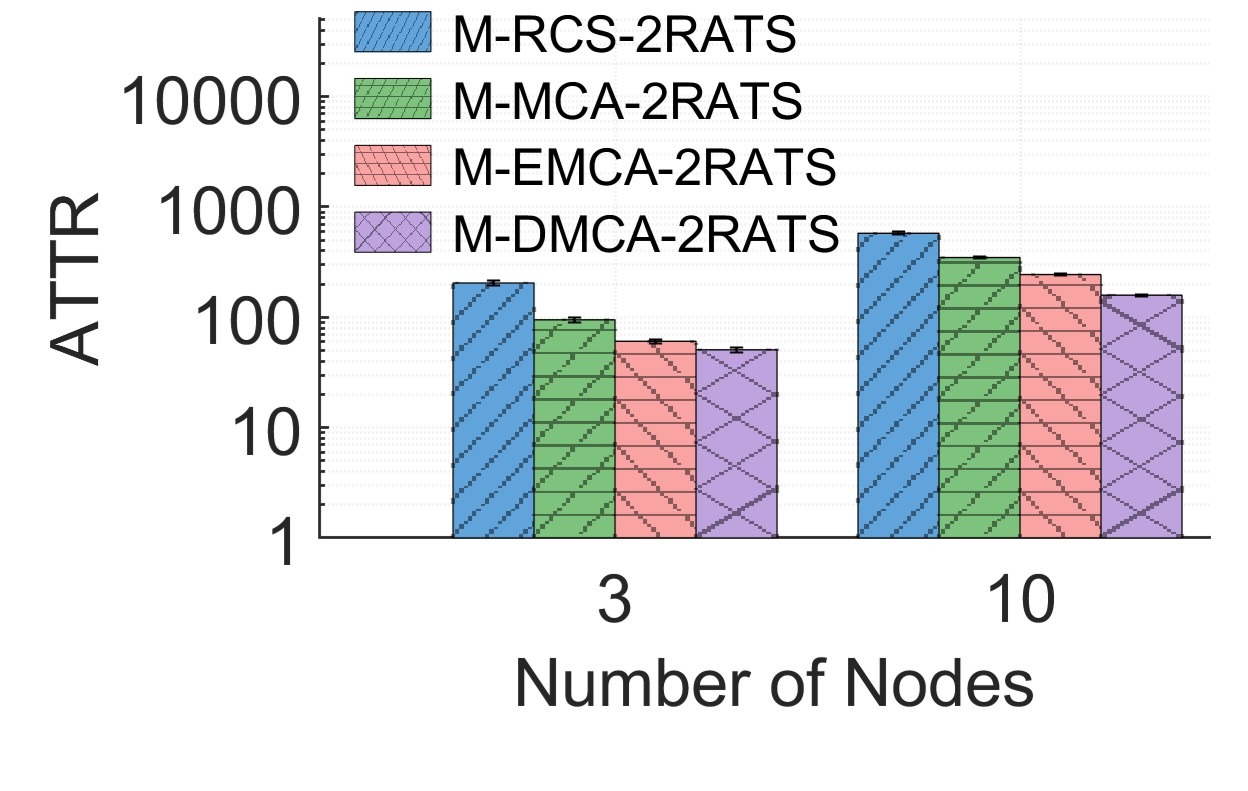}%
			};
		\end{tikzpicture}\\[-0.25em]
		{\footnotesize (b)}
	\end{minipage}
	\caption{Asym 10-CH with m=2 (a) 0\%PR (b) 85\%PR}\hfill
	\label{fig:R4}
\end{figure}

Figures~\ref{fig:R6}-\ref{fig:R4}, illustrates the effect of channel asymmetry and similarity ratio on ATTR. As m decreases from 9 to 2 (Fig.~\ref{fig:R6}a, ~\ref{fig:R5}a and ~\ref{fig:R4}a for 3WH), rendezvous time increases significantly due to reduced number of common channels. This effect is further intensified under 85\% PR activity (Fig.~\ref{fig:R6}b, ~\ref{fig:R5}b and ~\ref{fig:R4}b with 3WH), highlighting the challenge of multihop coordination in sparse spectrum conditions. M-DMCA still manages to improve the ATTR under these conditions by up to 49.6\% compared to all existing multihop rendezvous protocols.

\vspace*{-12.8pt}
\vspace*{-14.8pt}
\vspace*{-12.8pt}
\subsection{Scalability with larger channel set and number of nodes with mixed PR activity}
\begin{figure}[ht]
	\centering	
	% --- Left image (a) ---
	\begin{minipage}[b]{0.46\columnwidth}
		\centering
		\begin{tikzpicture}
			\node[draw, line width=0.5pt, inner sep=2pt] {%
				\includegraphics[width=1.36in]{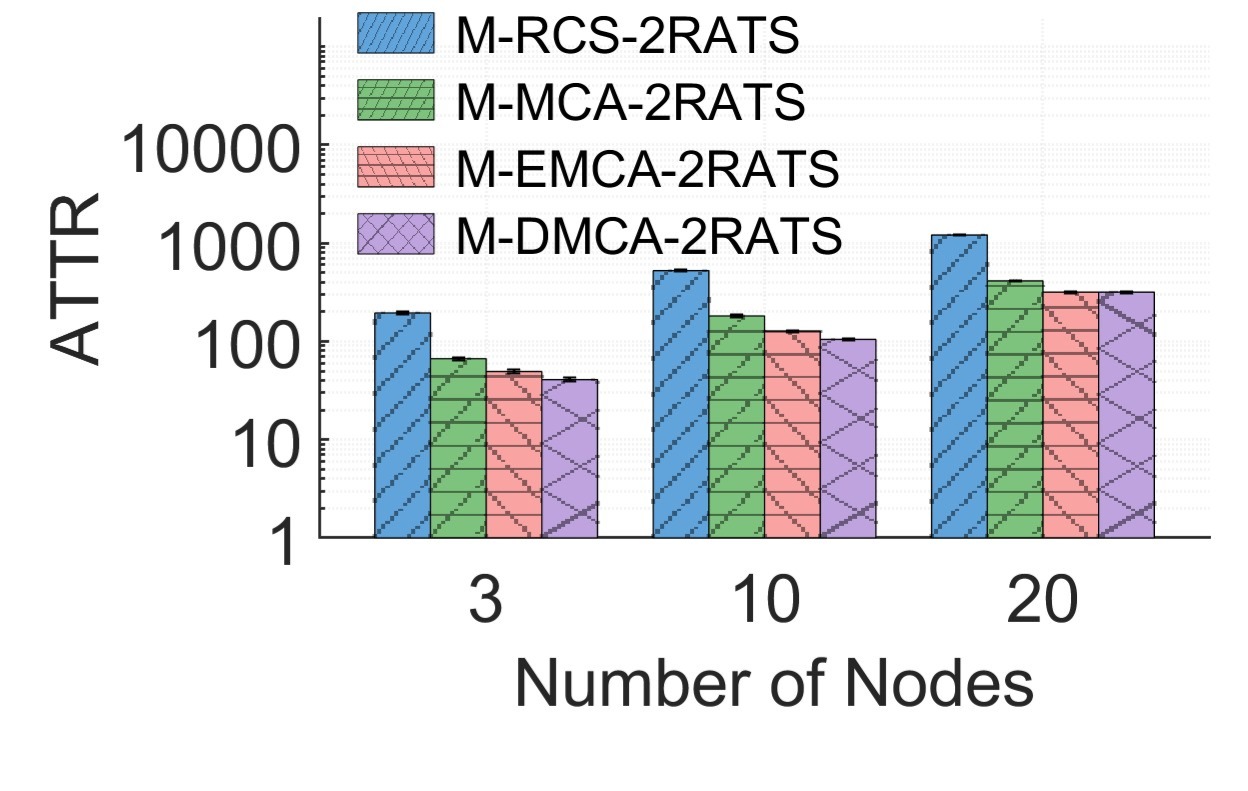}
			};
		\end{tikzpicture}\\[-0.25em]
		{\footnotesize (a)}
	\end{minipage}%
	%\hfill
	% --- Right image (b) ---
	\begin{minipage}[b]{0.46\columnwidth}
		\centering
		\begin{tikzpicture}
			\node[draw, line width=0.5pt, inner sep=2pt] {%
				\includegraphics[width=1.36in]{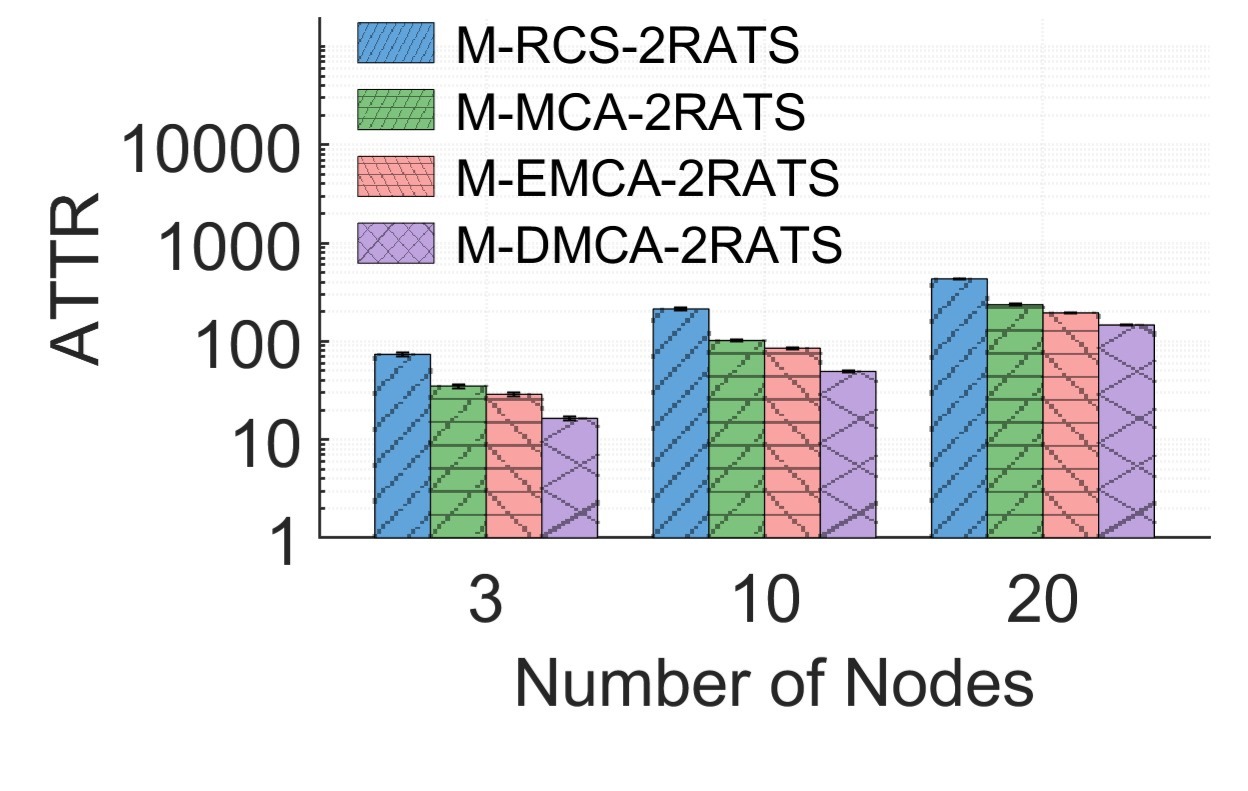}%
			};
		\end{tikzpicture}\\[-0.25em]
		{\footnotesize (b)}
	\end{minipage}
	\caption{Asym 20-CH with 0\%PR (a) m=2  (b) m=5}\hfill
	\label{fig:R78a}
\end{figure}

\begin{figure}[ht]
	\centering	
	% --- Left image (a) ---
	\begin{minipage}[b]{0.46\columnwidth}
		\centering
		\begin{tikzpicture}
			\node[draw, line width=0.5pt, inner sep=2pt] {%
				\includegraphics[width=1.36in]{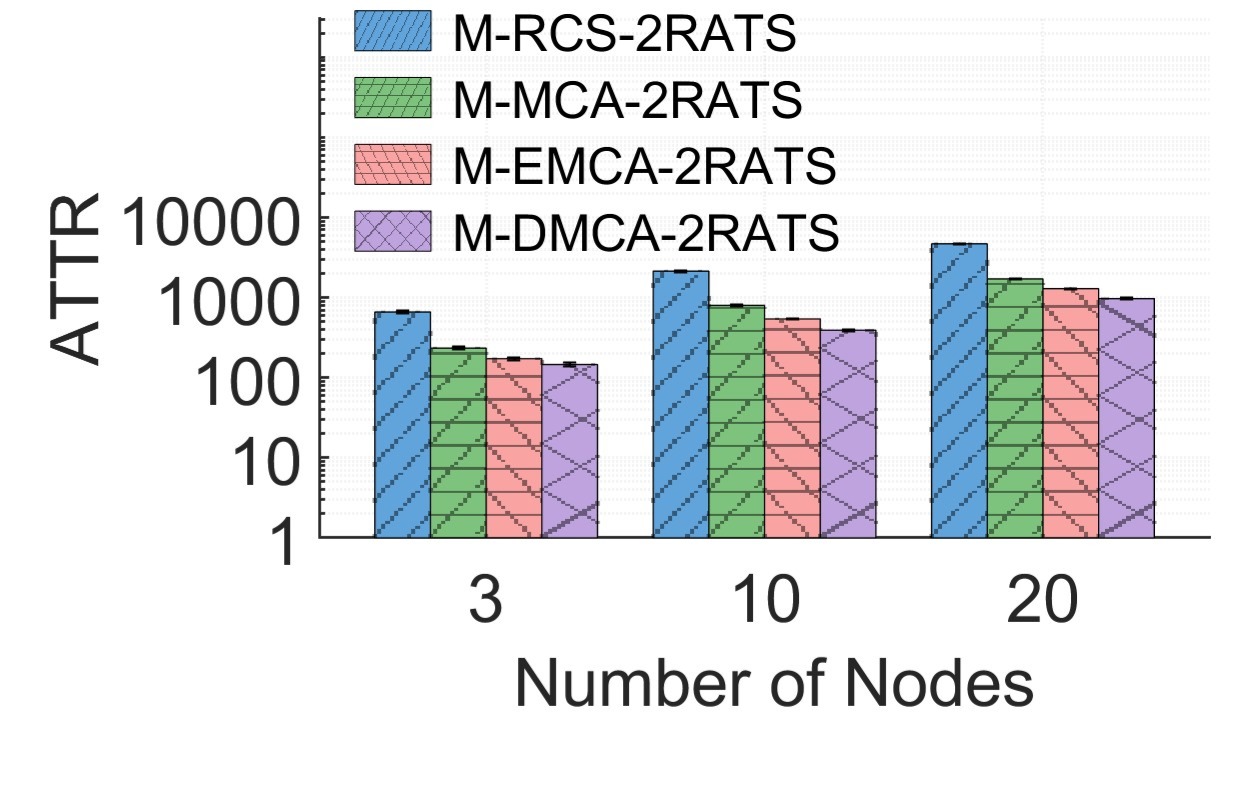}
			};
		\end{tikzpicture}\\[-0.25em]
		{\footnotesize (a)}
	\end{minipage}%
	%\hfill
	% --- Right image (b) ---
	\begin{minipage}[b]{0.46\columnwidth}
		\centering
		\begin{tikzpicture}
			\node[draw, line width=0.5pt, inner sep=2pt] {%
				\includegraphics[width=1.36in]{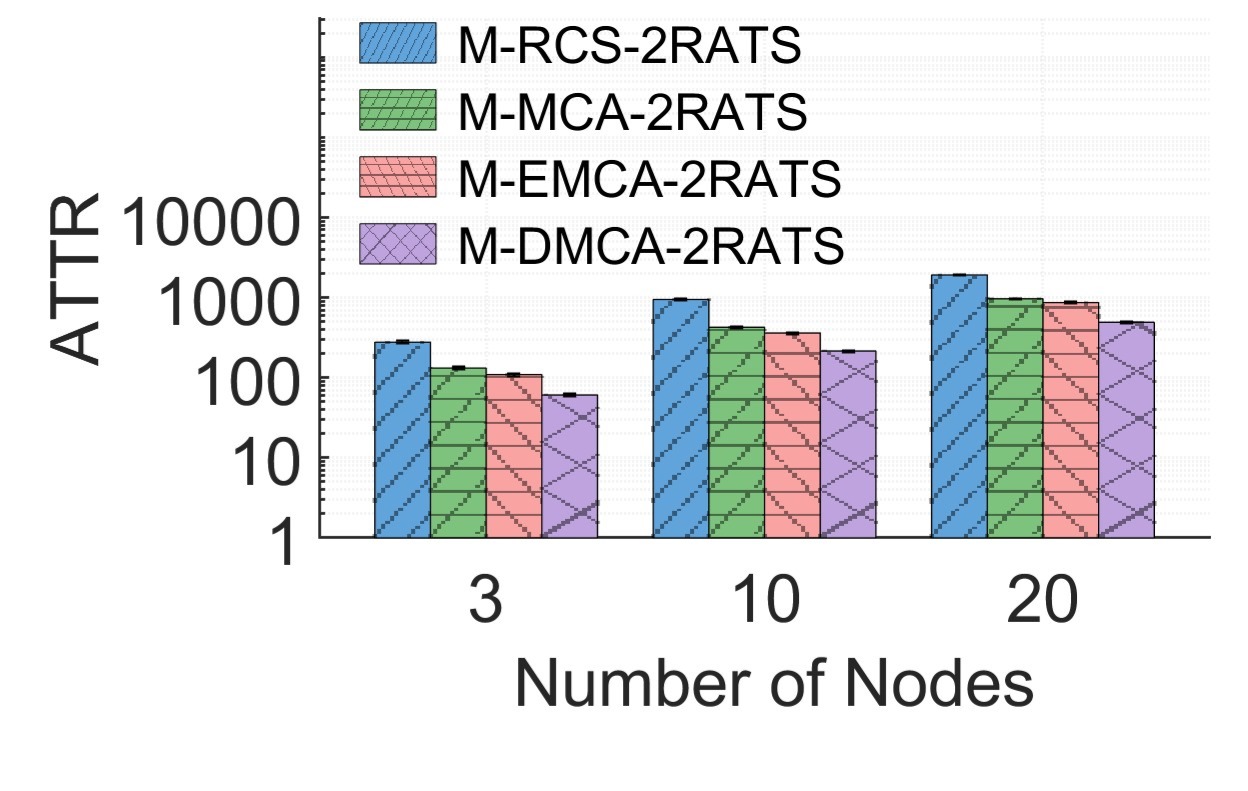}%
			};
		\end{tikzpicture}\\[-0.25em]
		{\footnotesize (b)}
	\end{minipage}		
	\caption{Asym 20-CH with 85\%PR (a) m=2  (b) m=5}\hfill
	\label{fig:R78b}
\end{figure}

\begin{figure}[ht]
	\centering	
	% --- Left image (a) ---
	\begin{minipage}[b]{0.46\columnwidth}
		\centering
		\begin{tikzpicture}
			\node[draw, line width=0.5pt, inner sep=2pt] {%
				\includegraphics[width=1.36in]{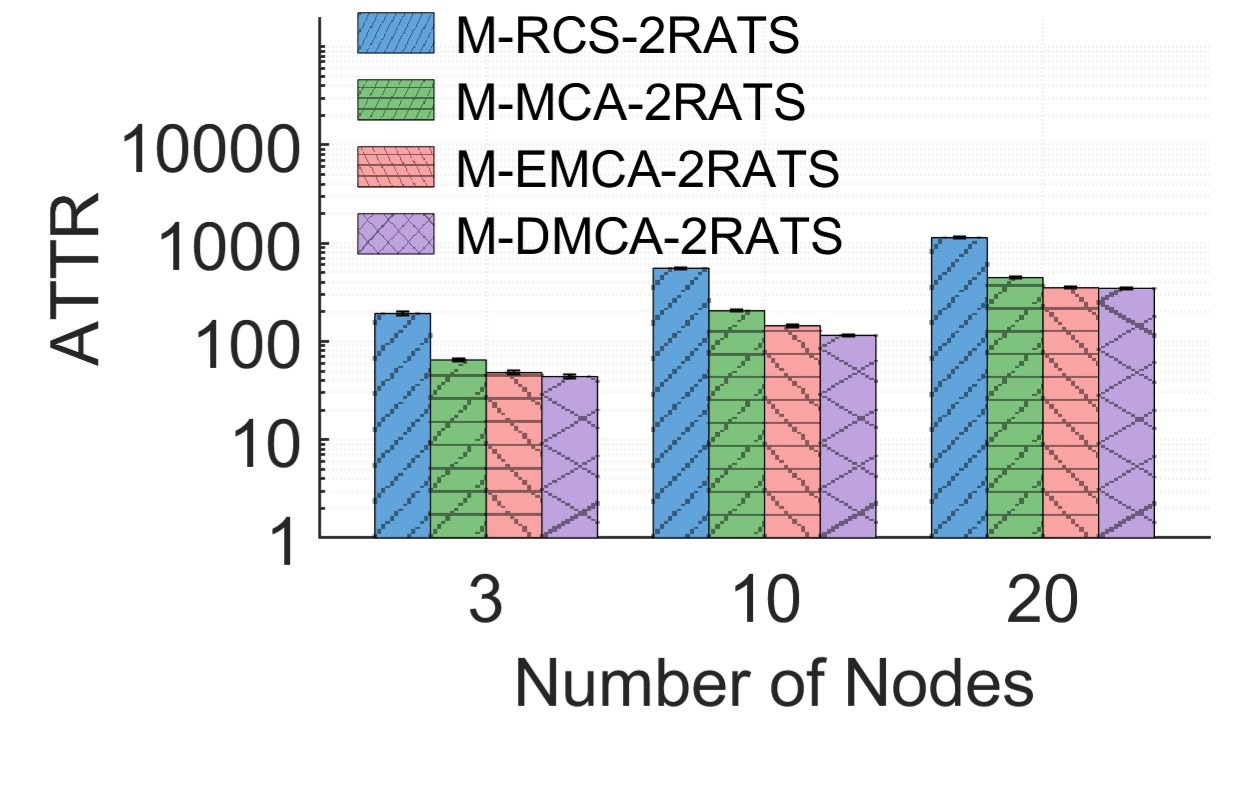}
			};
		\end{tikzpicture}\\[-0.25em]
		{\footnotesize (a)}
	\end{minipage}%
	% --- Right image (b) ---
	\begin{minipage}[b]{0.46\columnwidth}
		\centering
		\begin{tikzpicture}
			\node[draw, line width=0.5pt, inner sep=2pt] {%
				\includegraphics[width=1.36in]{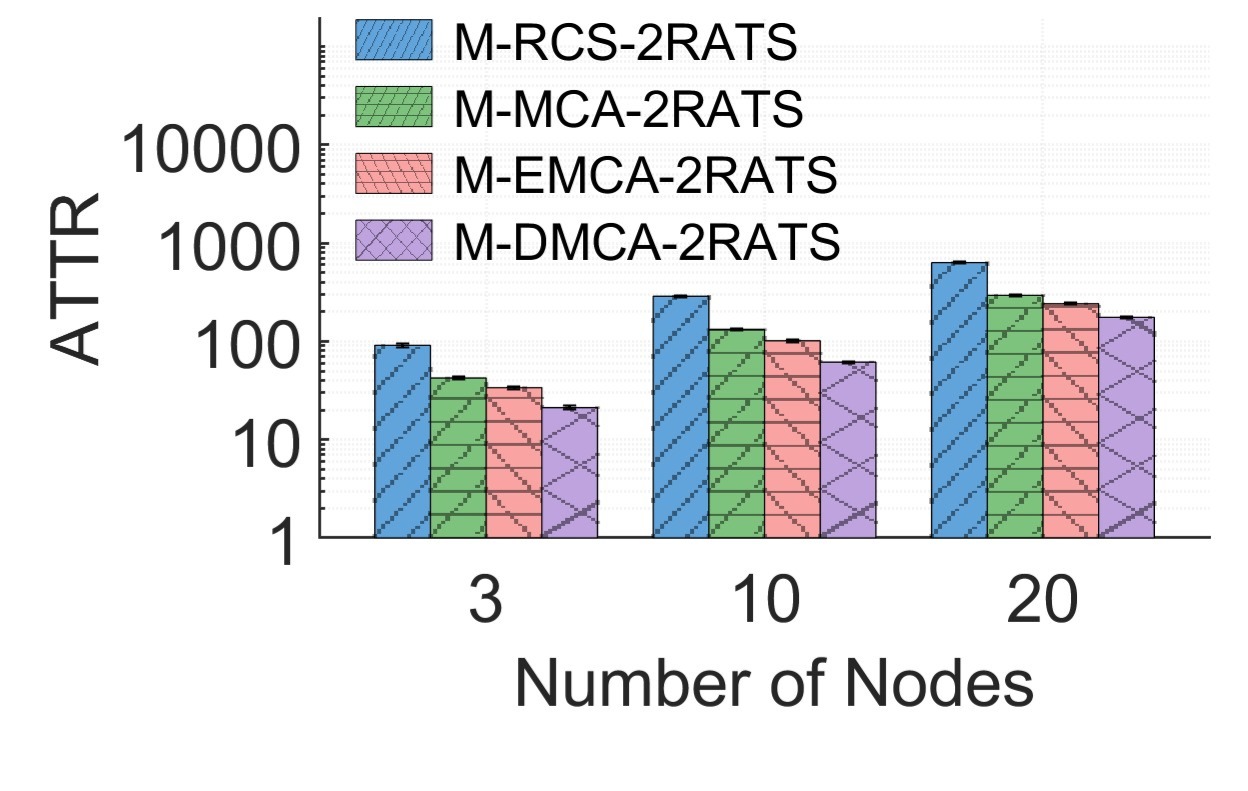}%
			};
		\end{tikzpicture}\\[-0.25em]
		{\footnotesize (b)}
	\end{minipage}
	
	\caption{Asym 20-CH with Mix PR (a) m=2  (b) m=5}\hfill
	\label{fig:R78c}
\end{figure}

Figures~\ref{fig:R78a}-\ref{fig:R78c} illustrates the performance of M-DMCA under 20 asymmetric channels (with m=2 and m=5) for scenarios involving 3, 10, and 20 nodes. As the number of channels increases, the rendezvous time increases significantly, as observed in Fig.~\ref{fig:R4} and~\ref{fig:R78a}. However, M-DMCA consistently maintains lower ATTR values compared to existing protocols. A similar trend is observed when the number of nodes increases from 10 to 20, as shown in Fig.~\ref{fig:R78a}. When PR activity increases to 85\% in Fig.~\ref{fig:R78b}, the rendezvous time becomes dramatically higher compared to the 0\% PR activity case shown in Fig.~\ref{fig:R78a}. This is due to the compounded effect of a larger number of asymmetric channels and increased node density with a similarity ratio of m=2, representing a worst-case scenario. Despite these conditions and sparse network situations, M-DMCA’s dual-channel selection far reduces 24\% ATTR compared to its closest M-EMCA rendezvous protocol which is better than all traditional rendezvous protocols.
Furthermore, when mixed PR activity is introduced to emulate unpredictable spectrum availability in critical scenarios, ATTR increases, as shown in Fig.~\ref{fig:R78c}. Although higher PR activity generally leads to increased ATTR, the mixed PR model demonstrates that M-DMCA maintains stable rendezvous performance despite unpredictable spectrum conditions, validating its suitability for disaster-response networks.

\vspace*{-12.8pt}
\vspace*{-12.8pt}
\vspace*{-12.8pt}
\section{Conclusion}
This paper presented M-DMCA, a multihop dual modular clock–based rendezvous protocol for cognitive radio–enabled emergency response networks. By supporting dual-channel selection and incorporating a three-way handshake, M-DMCA significantly reduces rendezvous time under symmetric and asymmetric channel, varying channel similarity ratios, mixed and high primary radio activity. Extensive NS-3 simulations demonstrate that the proposed approach outperforms conventional multihop rendezvous schemes, particularly for closest rendezvous protocol M-EMCA in challenging scenarios with limited common channels and dynamic spectrum availability conditions by upto 24\%. The approach also reduces communication overhead per successful rendezvous (PPR), highlighting the efficiency benefit of the three-way handshake.The current study assumes a known number of nodes, which is practical for many coordinated deployments. As future work, we plan to extend M-DMCA to support large channel sets incorporating unknown number of nodes and mobility, hence enhancing its capacity to autonomously address the challenges of more complex emergency response network.	  

%As future work, we plan to extend M-DMCA to support scenarios with an unknown and dynamically varying number of nodes, further enhancing its applicability to large-scale and ad hoc disaster environments.	       

%\section*{Acknowledgment}This Research has been supported by the Atlantic Technological University, Ireland under the Postgraduate Research Training Program in Modeling and Computation for Health and Society (MOCHAS PRTP).

\end{document}